\documentclass[
 reprint,
 amsmath,
 amssymb,
 aps,
amssymb]{revtex4-1}
\usepackage{amssymb}
\usepackage{graphicx}
\usepackage{dcolumn}
\usepackage{bm}%
\usepackage{gensymb}

\usepackage[dvipsnames]{xcolor}
\usepackage{ragged2e}

\renewcommand{\thefigure}{\textbf{\arabic{figure}}}

\begin{document}

\title{Hot Brownian Motion of thermoresponsive microgels in optical tweezers shows discontinuous volume phase transition and bistability}

\author{Miguel Angel Fernandez-Rodriguez$^{1,2*}$}
\author{Sergio Orozco-Barrera$^{1}$}
\author{Wei Sun$^{1,3}$}
\author{Francisco Gámez$^{1,4}$}
\author{Carlos Caro$^5$}
\author{María L. García-Martín$^{5,6}$}
\author{Ra\'ul Alberto Rica$^{1,*}$}

\affiliation{$^{1}$Universidad de Granada, Nanoparticles Trapping Laboratory, Department of Applied Physics and Research Unit “Modeling Nature” (MNat), Faculty of Sciences, Campus de Fuentenueva s/n, 18071 Granada, Spain
\\
$^{2}$Laboratory of Surface and Interface Physics, Department of Applied Physics, Faculty of Sciences, Universidad de Granada, Campus de Fuentenueva s/n, 18071 Granada, Spain
\\
$^3$Department of Physics, Yanshan University, Qinhuangdao, 066004, China
\\
$^4$Department of Physical Chemistry, Faculty of Chemical Sciences, Complutense University of Madrid, 28040 Madrid, Spain
\\
$^5$Instituto de Investigación Biomédica de Málaga y Plataforma en Nanomedicina (IBIMA Plataforma BIONAND), C/ Severo Ochoa, 35, 29590 Málaga, Spain
\\
$^6$Biomedical Research Networking Center in Bioengineering, Biomaterials \& Nanomedicine (CIBER-BBN), 28029 Madrid, Spain
\\
$^*$To whom correspondence should be addressed: E-mail:   mafernandez@ugr.es (MAFR); rul@ugr.es (RAR).}

\begin{abstract}
Microgels are soft microparticles that often exhibit thermoresponsiveness and feature a transformation at a critical temperature, referred to as the volume phase transition temperature. The question of whether this transformation occurs as a smooth or as a discontinuous one is still a matter of debate. This question can be addressed by studying individual microgels trapped in optical tweezers. For this aim, composite particles were obtained by decorating pNIPAM microgels with iron oxide nanocubes. These composites become self-heating when illuminated by the infrared trapping laser, featuring Hot Brownian Motion within the trap. Above a certain laser power, a single decorated microgel features a volume phase transition that is discontinuous, while the usual continuous sigmoidal-like dependence is recovered after averaging over different microgels. The collective sigmoidal behavior enables the application of a power-to-temperature calibration and provides the effective drag coefficient of the self-heating microgels, thus establishing these composite particles as potential micro-thermometers and micro-heaters. Moreover, the self-heating microgels also exhibit an unexpected and intriguing bistability behavior above the critical temperature, probably due to partial collapses of the microgel. These results set the stage for further studies and the development of applications based on the Hot Brownian Motion of soft particles. 
\end{abstract}

\maketitle

\date{\today}
\justifying

\textbf{Introduction.}
Microgels are soft systems comprised of crosslinked hydrogels that often exhibit thermoresponsiveness and collapse above a volume phase transition temperature ($VPTT$) \cite{Camerin2019, Zero}. The proximity of the $VPTT$ to physiological temperatures (e.g. around 32 $\degree$C for pNIPAM) together with their capability to carry cargo make them very well suited to develop interesting applications, including their use as drug-delivery carriers \cite{microgelSmall2022, BiomedicalApplications2018}, their potential applicability in the exploration of synthetic cell research \cite{ArtificialCell2020} and, more generally, to manufacture thermoresponsive surfaces and materials \cite{MicrogelSurfaceSmall2019,RobotGrip2019}. 

There are still open questions regarding the thermoresponsiveness of microgels, e.g., how the mechanical properties change across the $VPTT$ \cite{AFM2008}, or why there are multiple values of $VPTT$ reported for pNIPAM in the literature \cite{Phase,predictVPTT}. A more fundamental question remains unsolved, namely, whether the transition, i.e., collapse/swelling upon heating/cooling above/below the $VPTT$, is a sudden discontinuous or a smooth continuous process. One of the more usual ways to characterize thermoresponsiveness is to measure the variation of the average hydrodynamic size of microgel suspensions with temperature by dynamic light scattering (DLS). This variation is typically reported to follow a sigmoidal smooth curve \cite{Camerin2019}. Only recently, some works dealt with the characterization of the $VPTT$ transition of thermoresponsive microgels by optical tweezers, i.e., measuring properties in a one-microgel-at-a-time basis while externally heated \cite{Tata2011, Tata2016_T}. The results were disparate and in apparent contradiction. While the first work reported a discontinuous transition \cite{Tata2011}, the latter accounted for a smooth sigmoidal one \cite{Tata2016_T}, but being unspecified if the sigmoidal behavior were obtained by averaging results over several microgels. Moreover, the same authors found a rather discontinuous transition for pH-responsive microgels upon pH changes \cite{Tata2016_pH}, so the available results are inconclusive.  

Alternative à la carte synthetic routes widen the applications spectra of microgels by, for instance, adding co-monomers during the synthesis able to induce pH-responsiveness \cite{Pich2007} or embedding \cite{Lu2022, Hormeno_Small} or decorating \cite{PichDecoration2008, Liz2015, LizSmall2007} the microgels with inorganic nanoparticles that transfer their properties to the microgel to some extent \cite{MicrogelDecoration2015}. Hence, rather than being externally heated, self-heating microgels owning an absorbing core \cite{Lu2022,Hormeno_Small} or decorated with absorbing nanoparticles \cite{Hormeno2011,xiao2022synergic} might be assembled. These approaches reported continuous transitions too. From fundamental grounds, by adding absorbing nanoparticles in optical tweezers, we enter the interesting domain of Hot Brownian Motion (HBM) \cite{Volpe2021, HeatEngines2020, HotRotation2012, HotBrownian2010}, an intrinsically out-of-equilibrium process in which a particle is hotter than the surrounding medium, and gradients of temperature and viscosity develop. In this situation, the characterization of the temperature and other properties of the self-heating trapped particle becomes very difficult \cite{BrownianThermometry2020, EffectiveT_HBM2018}, and exotic properties in both the particle and the surrounding liquid can be observed \cite{TransitionWater2020}.

In this work, we decorated thermoresponsive pNIPAM microgels with iron oxide nanocubes. The dynamics and thermal behavior of the optically trapped composite were evaluated with the aim of both responding to the fundamental question raised before on the nature of the thermoresponsive transition and providing new insights and useful routes for characterizing the HBM of absorbing particles under optical trapping conditions. We found unexpected phenomena with potential broad applications as the accurate and localized heating and micro-thermometry of biologically relevant systems, and, in parallel, very intriguing bistability that could serve as a unique playground for statistical physics \cite{ricci2017optically} and soft active systems \cite{Moncho2022}.

\textbf{Results and Discussion.}
The pNIPAM microgels used in this work were synthesized by precipitation polymerization with a 4.86 \%\textsc{m} crosslinking density, resulting in a more crosslinked core and a less crosslinked shell \cite{Camerin2019, Zero}. We characterized the hydrodynamic diameter $d_H$ of the microgel suspensions by DLS as a function of the temperature $T$ (see \textbf{Figure 1a}). The results followed the usual smooth transition that can be fitted by the sigmoidal function \cite{Lu2022,Hormeno_Small}:

\begin{equation}
d_H(T)=d_{H,collapsed}+\frac{d_{H,swollen}-d_{H,collapsed}}{1+\exp\left(\frac{T-VPTT}{\tau}\right)}
\label{eq1}
\end{equation}

where the fitting parameters are the hydrodynamic diameters in the swollen ($d_{H,swollen}=911\pm14$ nm) and collapsed ($d_{H,collapsed}=556\pm13$ nm) states, the decay coefficient $\tau= 3.0 \pm 0.5$ K, and the $VPTT= 303.9 \pm 0.6$ K. In order to characterize the microgel morphology in detail, we deposited a monolayer at the water/air interface at a surface pressure of $\Pi$=5 mN$\cdot$m$^{-1}$, and transferred it onto a flat silicon substrate to measure their height by Atomic Force Microscopy (AFM, see \textbf{Figure S1a} and the \textit{Experimental Section} for further details). This measurement allowed us to characterize the individual topology of the microgels, and the core-shell morphology was reflected in the Gaussian profile exhibited in the AFM images (see \textbf{Figure S1b}). This property is ascribed to the microgels expansion at the interface caused by the interfacial tension and only counterbalanced by the internal elasticity, which is proportional to the crosslinking density\cite{Camerin2019}. In general, the size of the microgels at an interface is different from their size in bulk.

\begin{figure}
  \includegraphics[width=\linewidth]{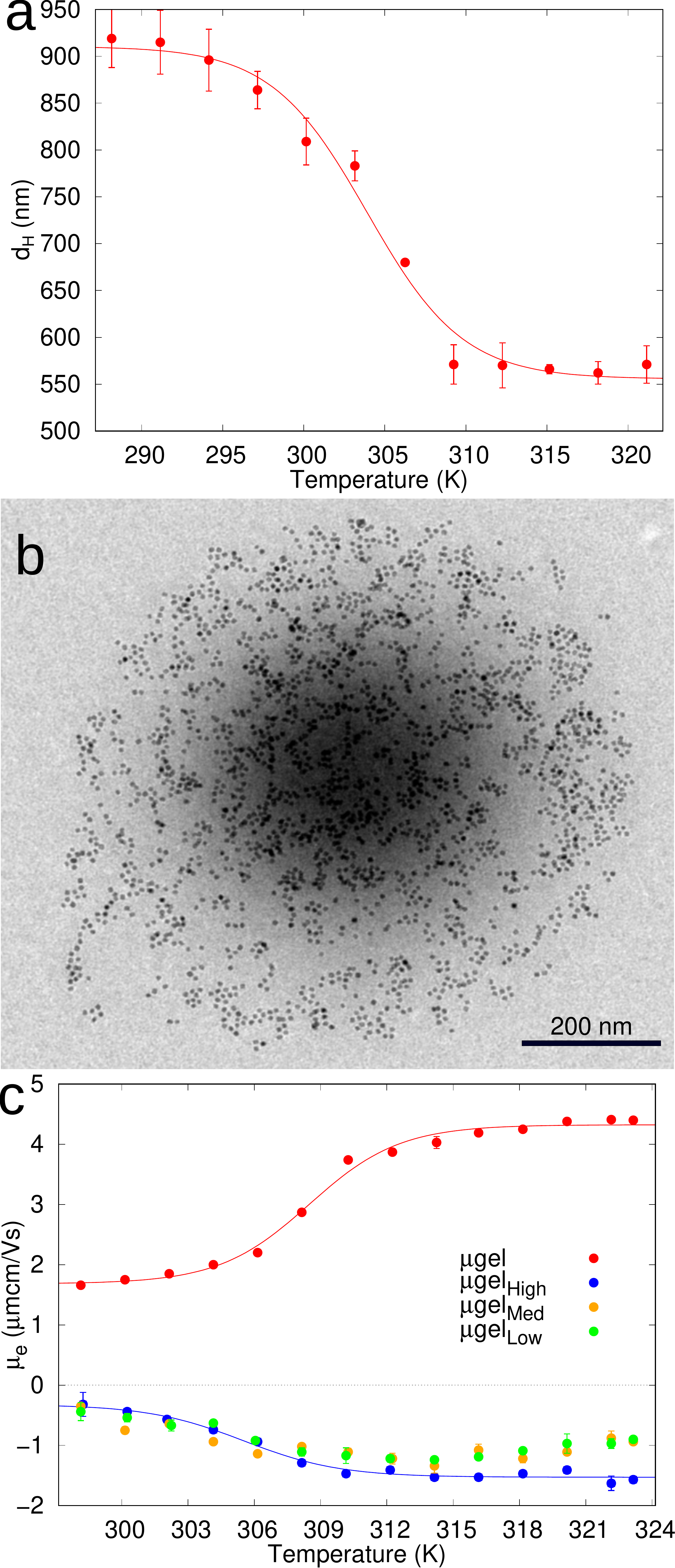}
  \caption{\textbf{(a)} Hydrodynamic diameter $d_H$ of the microgels measured by DLS against temperature. \textbf{(b)} TEM image of a $\mu gel_{High}$ decorated with the iron oxide nanocubes. \textbf{(c)} Electrophoretic mobility of the microgels against $T$ for the different prepared samples. The solid lines stand for sigmoidal fittings.}
  \label{fig:1}
\end{figure}

In order to make microgels moderately absorbing at the wavelength of our optical tweezers setup (wavelength of the trapping laser $\lambda=1064$ nm), we decorated them with iron oxide nanocubes of 12$\pm$2 nm of space diagonal length and 31$\pm$2 nm of $d_H$ with a polydispersity index of 0.30 (see the \textit{Experimental Section} and \textbf{Figure S1c-d} for further details about the size distribution and optical properties of the nanocubes). The electrostatically driven decoration was possible thanks to the positive charge of the microgels (see the red curve in \textbf{Figure 1d}) and the negative charge of the iron oxide nanocubes ($\zeta$-potential=--18$\pm$2 mV). The nanocubes got stuck on the polymer matrix, leading to composite particles (see \textbf{Figure 1b}) that are both thermoresponsive and self heat upon external illumination, as we demonstrate below. We tested three different relative fractions of nanocubes per microgel. For this purpose, we produced three samples labeled $\mu gel_{High}$, $\mu gel_{Med}$, and $\mu gel_{Low}$, in which we used a given amount of iron oxide nanocubes, half, and a quarter, respectively (see the \textit{Experimental Section} for further details). The charge of the composite was fully inverted with respect to the bare microgels in all cases (see \textbf{Figure \ref{fig:1}c}). This fact might point out specific interactions that not only induce electrostatic screening but a charge inversion \cite{moncho2014ion}. Despite this fact, the microgels preserved in all cases their thermoresponsiveness upon heating the whole dispersion, as it is shown in Figure \ref{fig:1}c. In this case, we can see that the charges per unit area, and hence the electrophoretic mobility $\mu_e$, increase in absolute value upon temperature-induced collapse. As compared to the bare microgels, the hydrodynamic diameter of the ones decorated with the highest amount of nanocubes ($\mu gel_{High}$) were reduced to $d_{H,swollen}=770\pm240$ nm and $d_{H,collapsed}=366\pm76$ nm, where the deviations account for the full width at half maximum (FWHM) of the corresponding measured distributions of $d_H$ (see \textbf{Figure S1e}). The widening in the distribution of $d_H$ might result from differences in the decoration with iron oxide nanocubes. Since the values of $\mu_e$ were similar in all decorated microgels, we considered that their $d_H$ were also similar.\\

The hydrodynamic diameter of \emph{individual} particles can be characterized in our optical tweezers device, thus avoiding averaging effects \cite{tolic2006calibration}. As we show below, this averaging becomes crucial in the interpretation of the collapse of microgels. Therefore, we studied single microgels in the optical tweezers device after dilution of the microgel dispersions in MilliQ water to a concentration down to a few particles per cm$^3$, in order to ensure that a single microgel was trapped at a time (see the scheme in \textbf{Figure 2a}). The evaluation of the diffusion coefficient $D=k_BT/\gamma$ of trapped particles is a well-established technique \cite{tolic2006calibration,Volpe_book}, where $k_B$ is the Boltzmann constant, $T$ is the temperature of the environment and $\gamma=3\pi\eta\cdot d_H$ is the friction coefficient of a sphere of hydrodynamic diameter $d_H$ in a liquid with viscosity $\eta$. A piezo stage that oscillates sinusoidally can be used to apply a known drag force on the trapped particle, providing direct measurements for the corner frequency of the particle in the trap $f_c$ and the diffusion coefficient (see the \textit{Experimental Section} and \textbf{Figure S2} for further details). We assumed that the light absorption by the bare microgels was negligible in the laser power $P$ range considered here, i.e., 8 to 351 mW at 1064 nm. This is reasonable since it is accepted that a 1064 nm laser heats water at a rate of 0.008 K$\cdot$mW$^{-1}$ when trapping a 500 nm silica nanoparticle \cite{Laser_heating}, corresponding to an increase of 3 K at our maximum $P$, which ensures that we do not reach the $VPPT$ from room temperature ($T=299\pm 2$K). Following this methodology, we evaluated the hydrodynamic diameter of 10 individual bare microgels at room temperature, obtaining an average value of $d_{H,single}=849\pm70$ nm. This result is in fair agreement with the average obtained for bare microgels from DLS measurements in the swollen state.

\begin{figure}[h!]
  \includegraphics[width=\linewidth]{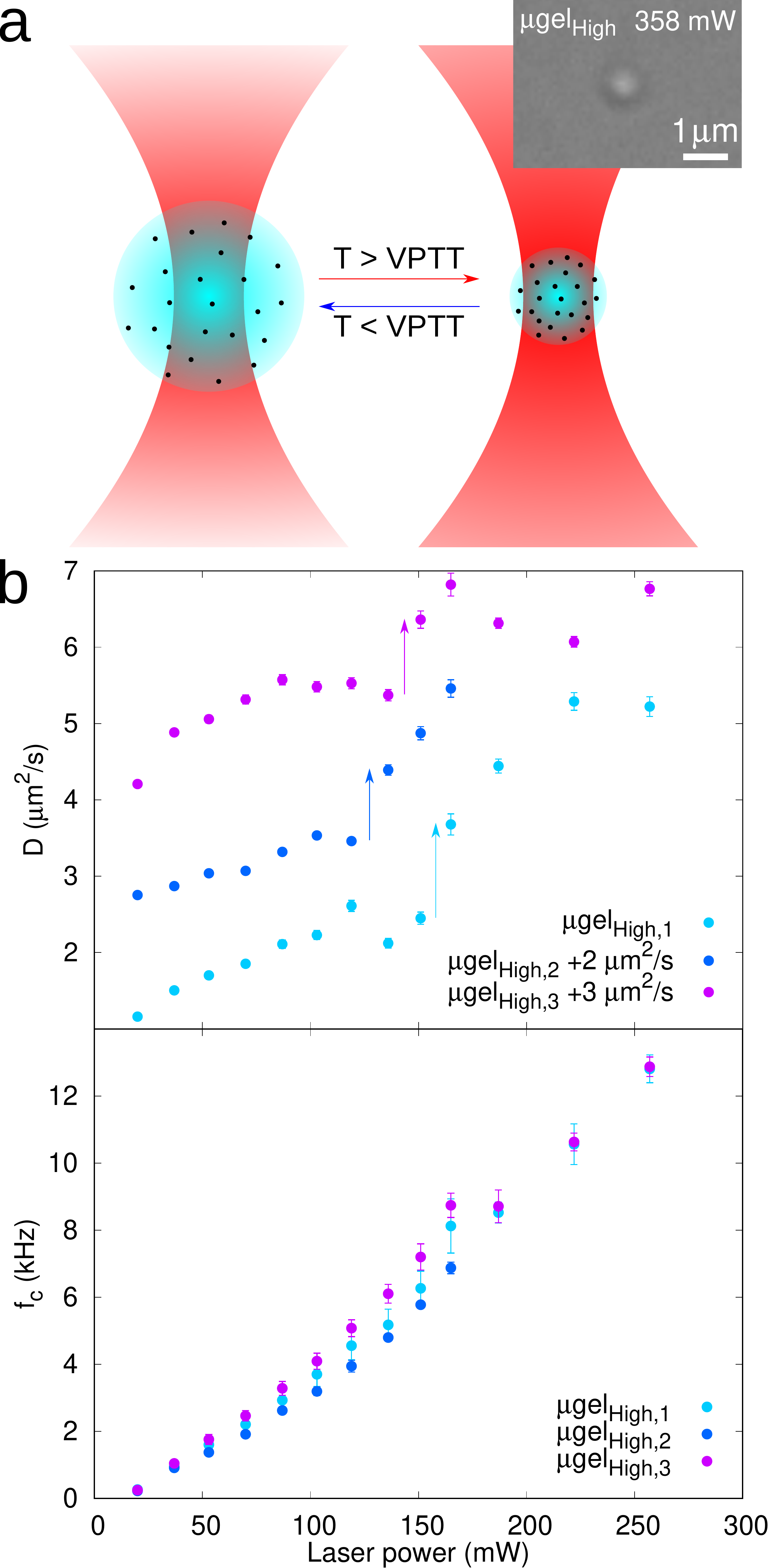}
  \caption{\textbf{(a)} Scheme of the optically trapped microgel decorated with iron oxide nanocubes below and above the $VPTT$. The inset shows a trapped microgel visualized from the optical camera. \textbf{(b)} Diffusivity $D$ of individual microgels and corner frequency $f_c$ as a function of the laser power $P$. The plots of $D$ are displaced vertically for each sample for the sake of clarity as detailed in the corresponding legends. Arrows depict the discontinuous transition for each microgel.}
  \label{fig:2}
\end{figure}

Decorated microgels trapped in optical tweezers presented clear signatures of heating due to the absorption of laser light, consisting in the observation of HBM and eventually collapsing upon increasing laser power. In order to demonstrate these effects, we resort to the evaluation of the diffusion coefficient of the decorated microgels, which can provide information about both effects. The measurement of the diffusion coefficient was performed following the same procedure described before for bare microgels. In \textbf{Figure \ref{fig:2}b}, we plotted $D$ as a function of $P$ for three individual self-heating microgel composites of the $\mu gel_{High}$ sample. The diffusion coefficient increases strongly with laser power, reaching a 5-fold increase at the maximum power we tested. There is a concomitant increasing trend of $f_c$ with $P$ which shows non-linear dependence, as expected for a sample that heats up upon laser power increase. It is also interesting to note that above the transition, the values of $f_c$ seem to coincide for the three microgels, pointing to an equivalent final state for all of them. A second striking feature in the dependence of the diffusion coefficient is a clear discontinuous transition at values $P\simeq 150$mW, slightly different for each microgel (see Figure \ref{fig:2}b). Regardless of the $P$ at which the transition was observed, the character is always discontinuous even when dozens of composites of each sample were evaluated. Interestingly, \textbf{Figure S3a-b} shows that the behavior of $D$ was reversible, as expected for the temperature-induced collapse of pNIPAM. 

To rule out the possibility that this discontinuous transition is an artifact of a particular way of measuring, complementary measurements of the electrophoretic mobility of the trapped composites $\mu_{e,trap}$  were performed. For this aim, we placed two parallel electrodes made of gold wires within the trap cell and subsequently applied a AC signal of $V_{pp}$=4 V at 20 kHz (see \textbf{Figure S3c}). Despite both measurements of $D$ and $\mu_{e,trap}$ being independent, discontinuous transitions at the same $P$ are derived for both parameters (see \textbf{Figure S3d}). To sum up, intriguingly, the one-by-one characterization of the temperature dependence of the diffusion coefficient and the electrophoretic mobility of optically trapped self-heating microgels leads to a discontinuous transition, which we proved to be reversible during a heating/cooling cycle. Previous results also showed discontinuous transitions for optically trapped non self-heating microgels \cite{Tata2011}, where the full cell was heated externally. For our self-heating microgels, we cannot distinguish if the dispersion in critical values of $P$ are due to slight heterogeneities in the crosslinking profiles or in the number of nanocubes embedded within the microgel. 

\begin{figure}
  \includegraphics[width=\linewidth]{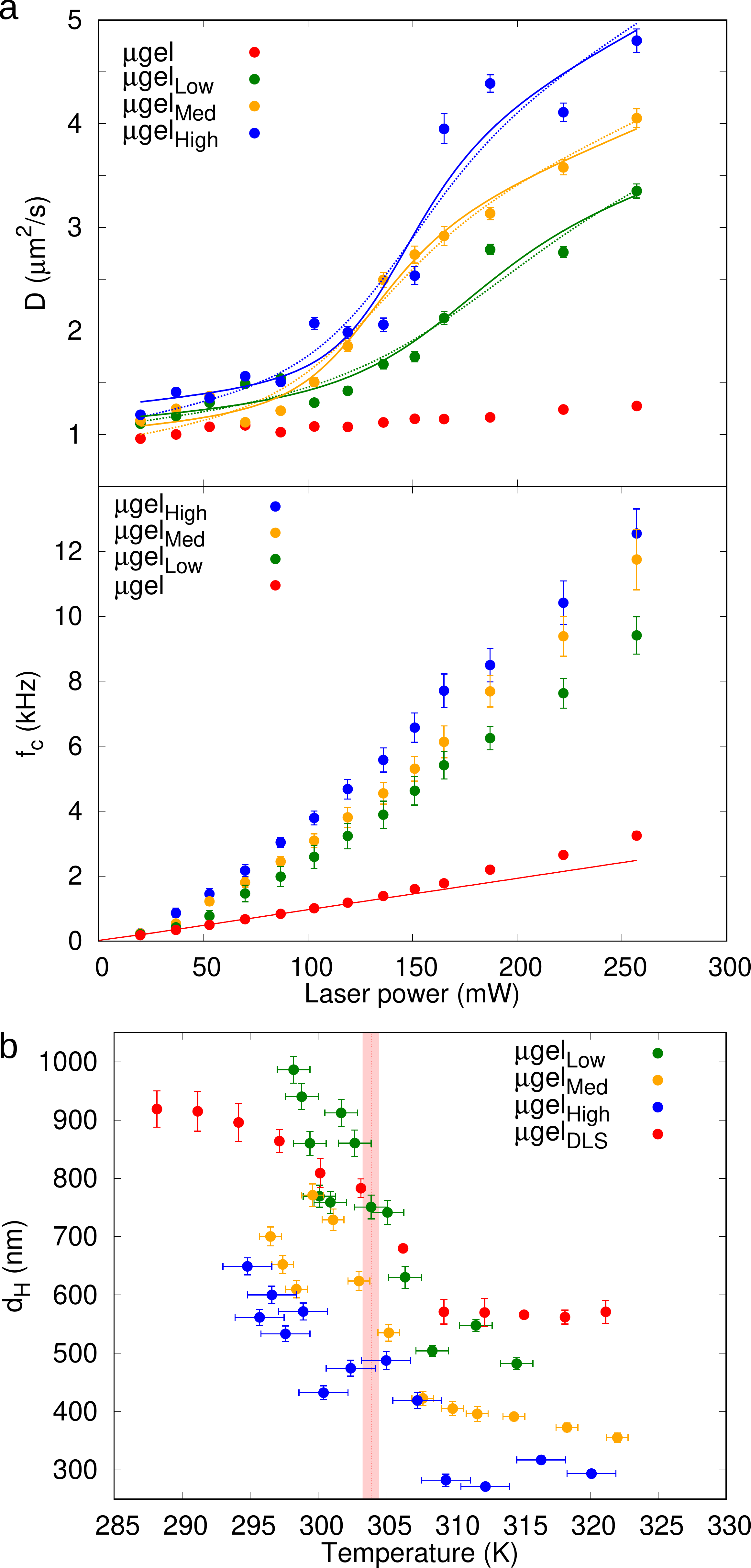}
  \caption{\textbf{(a)} Averaged $D$ and $f_c$ of 12 microgels for each sample. The dashed and solid lines are linear and Gaussian-like fittings, respectively, that allow to calibrate the $T$ corresponding to each $P$. The red line in the $f_c$ plot is the linear fitting y(x)=a$\cdot$x for the first points of the red data, to stress the deviation from linearity for higher $P$. \textbf{(b)} $d_H$ of the microgels as a function of $T$ by using the data fitted in (a) and Equation 1 and 2. The DLS size data from Figure 1b was added for comparison, and the red dashed line and shaded region correspond to the $VPTT$=303.9$\pm$0.6 K.}
  \label{fig:3}
\end{figure}

So far, we proved that there is a critical laser power above which a decorated microgel experiences a transformation. However, we still have not discussed whether this transition occurs at the $VPTT$, since no direct access to the local temperature $T$ of the self-heating microgel has been figured out yet. Moreover, a sigmoidal fitting does not render a good description for the transition of \emph{individual} self-heating microgels. The situation becomes different when the results are averaged over several self-heating microgels. In \textbf{Figure \ref{fig:3}a}, the variation of $D$ and $f_c$ against $P$ is plotted as obtained by averaging the power dependence over twelve individual microgels for each sample. Bare microgels feature a slight increase of $D$ with $P$ that might be related to the ability of the 1064 nm laser to heat the polymer and water, as discussed before. This was also evidenced by a departure from linearity in $f_c$ at higher $P$. In the case of self-heating microgels, averaging $D$ over several particles leads to a smooth sigmoidal-like transition, as it is typically observed in DLS measurements. It is hence tempting to argue that averaging over slightly different discontinuous transitions leads to a sigmoidal behaviour when averaging is performed either directly, like in DLS, or \textit{a posteriori}, as it is the case of our one-by-one experiments. Moreover, we see that the value of $f_c$  for the three samples of decorated microgels coincide at low powers with that of the bare ones, suggesting that the contribution to the polarizability of the iron oxide nanocubes within the polymer matrix is negligible and that all the observed effects could be attributed to self-heating.  

\begin{table}\centering
\caption{Fitting parameters of the linear $T(P)=a\cdot P+b$, and Gaussian-like $T(P)=a\cdot \exp\left(-2\left(\frac{d_H(T(P))\cdot r}{2w}\right)^2\right)P+b$ models, and drag-modification coefficient $\phi$, for the curves in Figure 2c. $\chi_\nu^2$ reported for each fitting.}
    \begin{tabular}{llllll}\hline
        Sample & Fit & $a$ (K$\cdot$mW$^{-1}$) & $b$ (K) & $\phi$ & $\chi_\nu^2$ \\\hline
        $\mu gel_{High}$ & lin. & $0.124\pm0.022$  & $288.4\pm2.6$ & $0.45\pm0.05$ & 0.13\\
        $\mu gel_{Med}$ & lin. & $0.118\pm0.012$  & $290.9\pm1.3$ & $0.57\pm0.03$ & 0.03\\
        $\mu gel_{Low}$ & lin. & $0.074\pm0.008$  & $292.6\pm1.8$ & $0.52\pm0.05$ & 0.03\\ 
        \hline
        $\mu gel_{High}$ & Gauss.  & $0.127\pm0.021$  & $293.8\pm1.8$ & $0.45\pm0.05$ & 0.11\\
        $\mu gel_{Med}$ & Gauss.  & $0.128\pm0.011$  & $295.5\pm0.8$ & $0.59\pm0.03$ & 0.02\\
        $\mu gel_{Low}$ & Gauss. & $0.085\pm0.010$  & $297.5\pm1.2$ & $0.58\pm0.05$ & 0.03\\\hline
    \end{tabular}  
\end{table}

Let us now develop a simple model that captures the main features of the observed behavior after averaging over experiments. The model should include both HBM due to the fact that the decorated microgels are hotter that the surrounding liquid, and the eventual collapse above the $VPTT$. We shall notice that a model for the HBM of soft particles is lacking, and beyond the scope of the present work. We will resort to a minimal model that allows us to estimate the transition temperature, in order to compare it with the reported values for $VPTT$. For this aim, we consider that the diffusion coefficient of a decorated microgel has a complex dependence on the temperature since both the size of the decorated microgel and the viscosity of water depend on it. The available models developed for hard spheres consider that the diffusion coefficient of a hot Brownian particle can be described by modified temperature $T_{\rm HBM}$ and viscosity $\eta_{\rm HBM}$, i.e., $D_{\rm HBM}=k_BT_{\rm HBM}/3\pi\eta_{\rm HBM}d_H$, where $T_{\rm HBM}$ relates the temperature at the hydrodynamic surface of the particle $T_s$ with that in the bulk far from the hot particle, $T_0$ \cite{HotBrownian2010,rings2011theory}. However, an alternative description is needed for our experiments, since the meaning of $T_s$ is lost in the case of microgels, and we need to account for its internal temperature, which will be non-uniform. Since no such model is available, we propose here a phenomenological model to capture both the HBM of a soft particle (SHBM) and the temperature-induced collapse:  

\begin{equation}
D_{\rm SHBM}(T, \phi)=\frac{k_B T}{\phi\cdot3\pi\eta(T) d_H(T)},
\label{eq:2}
\end{equation}

where the temperature is assumed to be the average internal temperature of the microgel and the factor $\phi$ accounts for the modified drag due to the gradient of temperature around the microgel and the concomitant viscosity one \cite{HotBrownian2010,Oppenheimer2016,BrownianThermometry2020}. We propose to combine the sigmoidal function in Equation \ref{eq1} with Equation \ref{eq:2} to fit our experimental data. Since the temperature dependence of the viscosity $\eta$ of water is well known \cite{eta} (see \textbf{Figure S4a}), we can use the measured values $d_{H,swollen}$ and $d_{H,collapsed}$ with DLS for the decorated microgels to obtain a non-linear expression of $D$ that increases with $T$ as shown in \textbf{Figure S4b}.

The missing ingredient that allows us to fit the data in Figure \ref{fig:3}a is the relation between the laser power and the effective temperature. The temperature increase is due to the absorption of laser light by the iron oxide nanocubes decorating the microgel. Two different calibrations are presented thereupon. Our first approximation assumes a linear increase of $T$ with $P$ ($T(P)=a\cdot P+b$), as it has been considered in several previous works \cite{Hormeno_Small,rodriguez2018temperature,magnetite_absorption}. Using this relation together with Eqs. \ref{eq1} and \ref{eq:2}, we can fit the data in Figure \ref{fig:3}a, obtaining the values of $a$, $b$, and $\phi$ presented in \textbf{Table 1} and the dashed lines in Figure \ref{fig:3}a. The first test of consistency is that $b$ should be the room temperature of the cell as $P=0$. Slightly lower values are in general obtained, featuring a very mild increase with the number of iron oxide nanocubes, and a similar increasing trend was obtained for the heating parameter $a$, compatible with reported values measured for NIR heating of multiple optically trapped iron oxide nanocubes of similar size at $P$=100 mW \cite{magnetite_absorption}. Our fitted $a$ contains the contribution from water heating due to the laser as discussed before, in the absence of self-heating nanoparticles, reported to be around 0.008 K$\cdot$mW$^{-1}$ for 500 nm silica nanoparticles \cite{Laser_heating}. Moreover, for more absorbing gold nanoparticles of 50 nm the heating coefficient was reported to be around 0.266 K$\cdot$mW$^{-1}$ for a single nanoparticle \cite{Gold}. Therefore, the iron oxide nanocubes seem to be a nice compromise to decorate the microgels as they become moderately self-heating, enabling us to observe the transition within the optical tweezers laser power range, and avoiding extreme phenomena such as cavitation. 

The linear dependence of $T$ with $P$ might be a good approximation for self-heating hard spheres, and below 320 K, where anomalous diffusion effects are reported \cite{TransitionWater2020}. However, it has been proven that the efficiency of heat absorption changes upon the collapse of a microgel shell over an inorganic absorbing core \cite{Lu2022}, and probably the effect is more dramatic in our case, where the absorbing nanoparticles decorate the shell of the microgel. In fact, upon collapse, the nanoparticles are brought closer to the center of the Gaussian profile of the laser beam \cite{rings2011theory}, where the light intensity is higher, and therefore the absorption should also increase. Moreover, it is reasonable to consider that the magnetite cubes will sit where the polymer density is higher. We can estimate this via a ratio between the collapsed and swollen states for the decorated microgels $r=d_{H,collapsed}/d_{H,swollen}\simeq0.5$. Then, $T$ would better be described using a Gaussian dependence of the form \\
$T(P)=a\cdot \exp\left(-2\left(\frac{d_H(T(P))\cdot r}{2w}\right)^2\right)P+b$, where $d_H$ still changes with $T$ according to Equation \ref{eq1} and $w=$280 nm is the beam radius of the trapping laser \cite{magnetite_absorption}. Since $d_H$ depends on the unknown value of $T$, the sigmoidal dependence of $T$ with $P$ was recursively solved using the results for the linear dependence to obtain the initial seed for $d_H(T(P))$. The sigmoidal fitting parameters are presented in Table 1 and as solid lines in Figure \ref{fig:3}a. It is apparent that the consistency test is better than when the linear dependence is assumed, with an average value of $\overline{b}$=296$\pm$2 K, in agreement with our room temperature. Nevertheless, the fittings did not significantly differ in practical terms as shown in Figure \ref{fig:3}a and \textbf{S5a}. For example, at $VPTT$=303.9 K, the difference of $P$ between the linear and Gaussian fittings is $|\Delta P|\simeq$4.5 mW, $\simeq$1.5 mW, and $\simeq$13.9 mW, for the $\mu gel_{High}$, $\mu gel_{Med}$, and $\mu gel_{Low}$, respectively. Consequently, we converted the $D(P)$ data from Figure \ref{fig:3} to $D(T)$ with both calibration approaches (see \textbf{Figure S5b})). By using the calibrated $D(T)$ we can retrieve a value of $d_H(T)$ by using Equation \ref{eq:2}, results that are presented in \textbf{Figure \ref{fig:3}b} and \textbf{S5c}. These plots need to be taken with a grain of salt, as they involve different simple and incomplete models relating $P$ with $T$. 

The first interesting feature we observe in Figure S5b is that the curves $D(T)$ show rather discontinuous transitions around the $VPTT$, also for the ensemble of microgels, i.e., after appropriate averaging. The maximum and minimum values of $d_H$ in Figure 3b show a decreasing trend with the content of iron oxide nanocubes, which showcases the ability of this technique to characterize the size of thermoresponsive soft particles when there is not enough sample to acquire DLS measurements. Nevertheless, we cannot capture the upper part of the sigmoidal curve as there is heating as soon as the particle is optically trapped. Furthermore, the Gaussian fit seems to provide better results, with better discrimination between samples and a value for $d_H$ above the $VPTT$ that remains rather constant, as shown in Figure \ref{fig:3}b and S5c. The results indicate that the decoration with iron oxide nanocubes does not seem to significantly alter the value of the $VPPT$, as also shown in Figure 1c from DLS measurements. It also gives value to our phenomenological model, as the transition is observed at the $VPPT$ and thus it seems reasonable to assume that the $T$ in Equation \ref{eq:2} is related to the real internal temperature of the microgel, and not only to the temperature at the hydrodynamic surface of the hot particle. Finally, we estimated the stiffness of the trap as a function of $P$ and $T$ as $\kappa = 2\pi f_c \gamma_{\rm SHBM} = 6\pi^2 f_c \phi \eta(T) d_H(T)$ for both calibrations (see \textbf{Figure S6}). There are mild changes in trend above the $VPPT$ that are clearer for the Gaussian-like calibration, and as discussed for $f_c$, $\kappa\neq0$ for $P=0$ mW which points out to the complex SBHM behavior.

The agreement of our observations with trapped microgels and the $VPTT$ reinforces the assumption that the free parameter $\phi$ accounts for HBM effects only related to drag. The average value of $\phi$ is 0.51$\pm$0.06 for the linear fitting (0.54$\pm$0.08 for the Gaussian fitting), which means that the effective $\gamma_{\rm SHBM}\simeq$ $\gamma$/2, i.e., a $\simeq$ 50 $\%$ smaller with respect to the drag of a sphere. This result benchmarks the effects of HBM on a soft particle and could be compared with the changes in drag coefficient due to the deformation of a soft spherical particle to a pancake shape, estimated to decrease $\gamma$ by $\simeq$ 60 $\%$ when it completely flattens \cite{Oblate2012}. This would not be supported by the optical images of the trapped microgel, whose size was observed to be compatible with the one from DLS, and would likely be in conflict with the collapse of the microgel that we observe. Summing up, it has been demonstrated that the self-heating microgels prepared here anew showed a discontinuous transition when isolated in an optical trap, while a sigmoidal-like continuous transition was obtained averaging over a dozen of experiments. In the latter situation, a fitting of $T$ at each laser power density can be established, enabling (i) their use as both local micro-heaters and micro-thermometers with potential applicability in biological studies, and (ii) providing a new way to measure the effective $\gamma_{\rm SHBM}$ that we expect to be helpful to delve into the HBM of both hard and soft particles from direct experimental data.

\begin{figure*}
  \includegraphics[width=\linewidth]{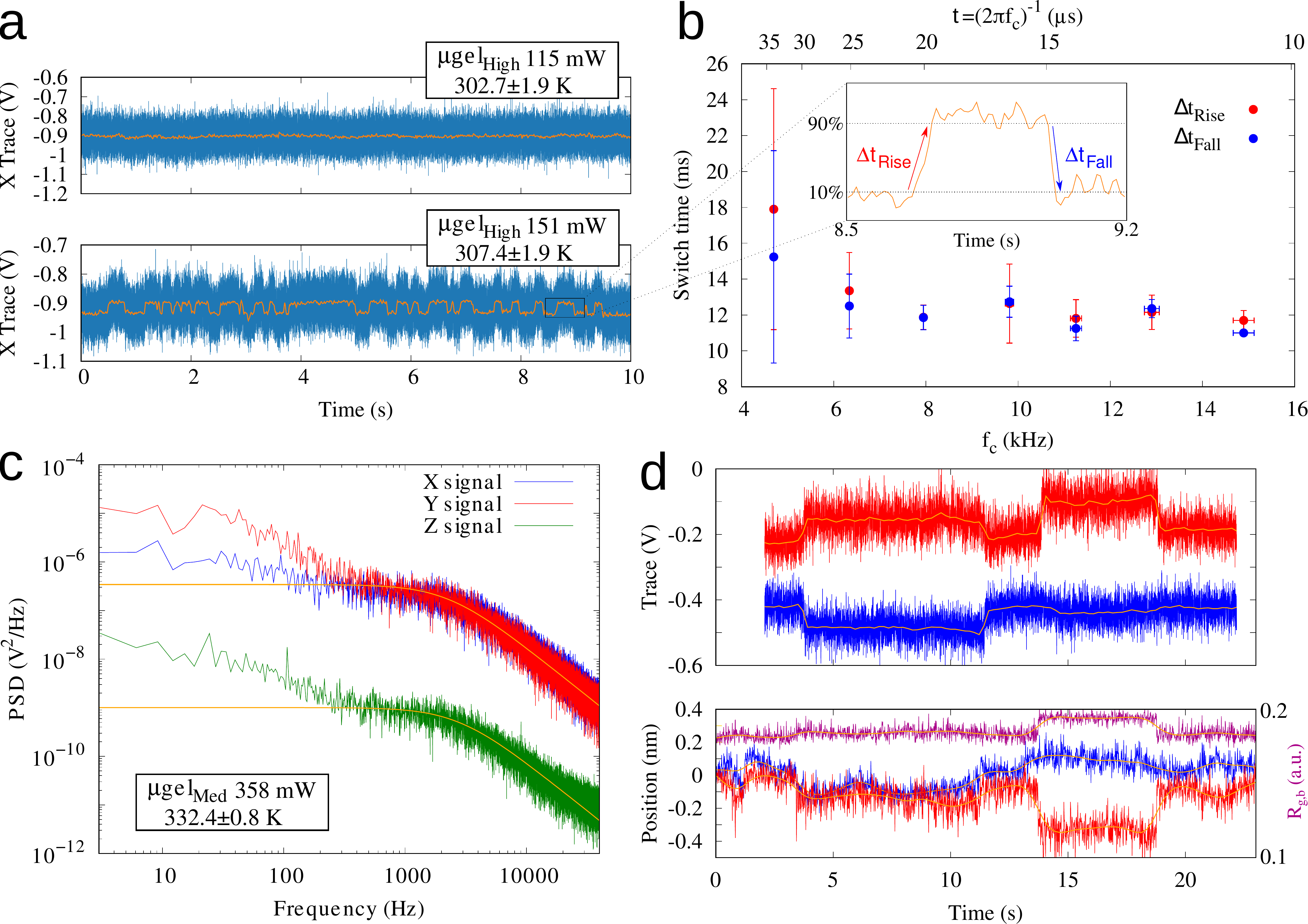}
  \caption{X raw traces for a $\mu gel_{High}$ in blue, and running average over 400 points in orange, below and above the $VPTT$.  \textbf{(b)} Switching times for the bistability, rise, and fall in the X trace as a function of $f_c$ and $t=(2\pi f_c)^{-1}$. \textbf{(c)} PSD of the X, Y, and Z traces for a $\mu gel_{Med}$ above the $VPTT$. The Gaussian fitting was used to estimate $T$ in (a-c). \textbf{(d)} Comparison between the features visible in the traces in volts (top) and by video tracking (bottom) of the X and Y centroid position, and the radius of gyration of the microgel brightness.}
  \label{fig:4}
\end{figure*}

Finally, we turn our attention back to the individual behaviour of decorated microgels to uncover an unexpected feature of their dynamics while trapped in optical tweezers. Taking a look at the individual traces we observed that, above the $VPTT=303.9\pm0.6$ K, a kind of bistability appeared in the time traces (see \textbf{Figure \ref{fig:4}a}), similar to the behavior reported for bistable trapping potentials \cite{rondin2017direct,zijlstra2020transition,bistable}. It is worth having in mind that the difference between the linear and Gaussian fittings at both 115 and 151 mW is $|\Delta T|<$ 0.8 K. Thus, we could make a reasonable claim that we were below and above the $VPTT$ for those laser powers, respectively. The bistability was typically observed in both X and Y signals of the photodiodes, which correspond with the trapping plane, being clearly correlated (see \textbf{Figure S7-8} for several time traces at different powers and their corresponding PSDs, respectively). Less clear was the presence of bistability in the Z signal, which corresponds with the direction of the laser beam. It is not clear if this is a fundamental property of the phenomenon or if this is an artifact of the detection along this axis, which is generally accepted to be less reliable \cite{tolic2006calibration}. We extracted the characteristic switching times as a function of $f_c$ (and thus the characteristic relaxation time in the trap $t=(2\pi f_c)^{-1}$), finding no clear dependence with it (see \textbf{Figure \ref{fig:4}b}). This is true even if the typical residence time clearly increases with $P$, and hence with $f_c$ (see Figure S7).

Interestingly, this behavior was accompanied by an increase in the plateau of the PSDs at low frequencies (see \textbf{Figure \ref{fig:4}c}). This feature was previously reported in microgels with absorbing inorganic nanoparticles in the shell \cite{Hormeno_Small}. We believe this might be a signature of self-heating in optical trapping, and it would be interesting to understand if the bistability appears also in their self-heating microgels. For comparison purposes, in \textbf{Figure S9a} we present the PSD for both a non-absorbing silica microparticle (1 $\mu$m SiO$_2$, Sigma Aldrich) and a self-heating hard microparticle (0.883 $\mu$m ProMag 1 with 26.5 $\,wt\%$ magnetite content, Bangs Laboratories). The latter displayed a rise in the  PSD at low frequencies. Nevertheless, in the case of a hard self-heating microparticle we did not observe bistability. 

In order to ensure that this effect is not an artifact from the detection with photodiodes, we recorded a movie with a CMOS camera of a microgel under bistability conditions and compared the traces from the photodiodes and the position and radius of gyration of the brightness $R_g$ after tracking the self-heating microgel position from the movie (see \textbf{Figure \ref{fig:4}d} and \textbf{Movie S1}). The same trends were observed in both the position and $R_g$. The jumps in the traces in volts correspond to very subtle movements of the microgel that are not visible in Movie S1 by the naked eye since they correspond to variations of the centroid position below $\simeq$50 nm. Furthermore, the PSD for the different steps of bistability in Figure \ref{fig:4}d were plotted in \textbf{Figure S9b}) and we found it to be similar to those PSDs in Figure \ref{fig:4}c. This fact discarded that the increase of the PSDs at low frequencies was only due to the bistability. 

We do not have enough evidence to elucidate the mechanism behind this bistability. We identified as possible hypotheses: i) displacements of the microgel within the trap, ii) spatial reconfigurations of the adsorbed iron oxide nanocubes within the soft microgel, iii) an alternation of the orientation between two birefringence axes of the collapsed microgel, and/or iv) partial collapses of the microgel, a phenomenon that it is currently under study \cite{2step,Monte2021,granules}. The first hypothesis would be supported by characteristic switching times of the order of the characteristic relaxation time in the trap $(2\pi f_c)^{-1}\simeq$ [0.01-0.05] ms, but we found them to be $\simeq$ 12 ms, independently of $P$ and hence of $f_c$ (see Figure \ref{fig:4}b). For the second hypothesis, we believe that the reconfiguration of the iron oxide nanocubes would lead to continuous changes rather than bistability,  since it is difficult to imagine the nanocubes jumping between two configurations within the microgel matrix. For the birefringence hypothesis, we estimated that the scattering force provided by the iron oxide nanocubes is twice the one necessary to deform the microgel by $\Delta V/V$ by 10\% (see \textbf{Figure S10}), which would support the idea of a microgel deformed to the point of exhibiting birefringence. Nevertheless, in this case, we would also expect a dependence of the switching time with $f_c$, as it would be related to the torque exerted by the linearly polarized laser, and hence should also decrease with $P$ \cite{HotRotation2012,li2016optical}, contrary to what we observe. The latter hypothesis on partial collapses might be tested if the changes in size were big enough to show differences in $f_c$, but when we checked the $f_c$ of the different segments of the bistable traces (see Figure S9b) they were not significantly different. Therefore, further studies are needed to elucidate the origin of this bistability. Regardless of the origin, this behavior constitutes an interesting model for statistical physics and poses a challenge for scientists in the search for answers in related fields \cite{Moncho2022}.

\textbf{Conclusions.}
By combining thermoresponsive pNIPAM microgels with iron oxide nanocubes, we obtained a self-heating soft microparticle system that was interrogated by optical tweezers. Both the diffusion coefficient $D$ and the corner frequency $f_c$ of the microgels were obtained under an assumption-free experimental approach. We found that single trapped self-heating microgels showed a discontinuous volume transition upon heating, while the usual sigmoidal-like transition was recovered when averaging over dozens of measurements. These results provide new insights into the discussion of the nature of the thermoresponsive transition of pNIPAM microgels. In particular, we provide experimental evidence that the sigmoidal-like thermal behaviour of self-heating microgels might come from collective effects due to subtle heterogeneity features in the size or chemical compositions of the microparticles. This evidence also explains the conclusions aroused when bulk techniques, such as DLS, are employed to monitor the $VPTT$. The collective sigmoidal-like behavior enabled us to model the dependence of $T$ on $P$, by either assuming a linear or a Gaussian-like dependence. This procedure turned the self-heating microgels into a dual system that simultaneously acts as both local micro-heaters and micro-thermometers. This feature is highly desired in nanomedicine and material sciences. At the same time, they provide a new way of characterizing the effective drag $\gamma_{\rm SHBM}$, a parameter that might be helpful for studying the Hot Brownian Motion of soft particles. Furthermore, we found an intriguing bistability in the time trace obtained above the $VPTT$ for single microgels that might be ascribed to partial collapses of the self-heating microgels, among other working hypotheses. 

Finally, we hope that the potential applications and fascinating bistability features of soft self-heating microgels demonstrated in this work shall spark the interest of a broad spectrum of the scientific community, from experimentalists focused on applied science to fundamental statistical physicists.

\section{MATERIALS AND METHODS}
\subsection{Synthesis and characterization of pNIPAM microgels}
The microgels were synthesized by precipitation polymerization as described in our previous works \cite{Zero}. First, we prepared MilliQ water at pH 4 by adding HCl with a glass pipette (Sigma Aldrich, ACS reagent 37 \%). Then, we added it (55 mL) to a round bottom glass flask and further added n-isopropylacrylamide (1.108 g, NIPAM, Sigma Aldrich, 99 $\%$) monomer and bis-acrylamide (0.077\,g, BIS, Sigma Aldrich, 99.5 $\%$) crosslinker, obtaining a crosslinking density of 4.86 $\%$\textsc{m}. Then, we gently shaked the flask until all the powder was dissolved in the water, and we placed a portion of this solution (5 mL) in a glass vial that we sealed with a rubber septum. Next, we added a teflon-coated magnetic stirrer and sealed the flask with another rubber septum, heating in an oil bath (80 $^\circ$C) under magnetic stirring and N$_2$ bubbling for 1 h to remove the oxygen in the flask. In parallel, we had the vial (5 mL) with the monomer and crosslinker under N$_2$ bubbling and we prepared another glass vial (6 mL) with pH 4 MilliQ water and added 2,2'-azobis(2-methylpropionamidine)dihydrochloride (0.088 g, V50, Acros, 98$\%$) initiator, and sealed the glass vial with another rubber septum, bubbling N$_2$ for 1 h. Next, we added the V50 initiator solution (2 mL) with a syringe and in a few minutes the solution turned white. In order to increase the size of the microgels, 15 min later we added more monomer+crosslinker solution (1.25 mL) from the separated glass vial and the V50 initiator (1 mL) every 10 min, for a total of 4 aliquots. Next, we kept the flask for 5 h at 80 $^\circ$C under N$_2$ atmosphere. The dispersion of microgels was then filtered with Whatmann paper (16$\,\mu$m pore size) and a glass funnel. In order to clean the sample from unreacted species we performed 5 cycles of centrifugation (39800 rcf) for 1 h each, removing the supernatant and redispersing the microgels in MilliQ water.
\vspace{0.5cm}

After this, we diluted it to 0.24 $\,wt\%$ with MilliQ water. In order to characterize the microgels, first we prepared a 4:1 microgel dispersion:isopropanol (Sigma Aldrich, HPLC grade) and added drops with a glass microsyringe to a water/air interface in a Langmuir trough (KSV NIMA, Biolin Scientific), where we previously immersed a silicon substrate (2 $\times$ 1 cm$^{2}$, $\left \langle 100 \right \rangle$ orientation, p-type, Boron doped, 1-10 $\Omega\cdot$cm, University Wafer) cut by laser (Laser E-20 SHG II, Rofin) and supported in a dipper. We added the dispersion at the interface until the surface pressure reached a value of $\Pi$=5 mN$\cdot$m$^{-1}$, and a Langmuir-Blodgett film was then deposited the by raising the dipper at 0.5 mm per min, keeping $\Pi$ constant with the movable barriers. After this, the monolayer was characterized by AFM (Dimension 3000) in tapping mode (Tap300Al-G cantilevers, 300 kHz, 40 N$\cdot$m$^{-1}$, BudgetSensors). We also characterized their electrophoretic mobility $\mu_e$ by Laser Doppler Micro-electrophoresis (ZetaSizer NanoZ, Malvern), and their $d_H$ by Dynamic Light Scattering with a capillary cell that allows to measure big nanoparticles (DLS, Zetasizer Ultra, Malvern).

\subsection{Synthesis of iron oxide nanocubes. }
Iron oleate precursor was synthesized as previously described in reference \cite{nanocubes}. Briefly, a mixture of iron chloride (10.8 g, 40 m\textsc{m}) and sodium oleate (36.5 g, 120 m\textsc{m}) were dissolved in ethanol (80 mL), distilled water (60 mL) and hexane (140 mL). The resulting solution was heated (60 $^\circ$C) and let for 4 h in a reflux of hexane and in inert atmosphere. At that time, the reaction was cooled down to room temperature and two phases can be distinguished: a lower aqueous phase and an upper organic phase with the iron oleate. The organic phase was washed 3 times with distilled water and the hexane was evaporated in the rotavapor. Once the precursor is synthesized, a thermal decomposition approach was selected for the synthesis of iron oxide nanoparticles. Following this method, iron oleate (1 g) and oleic acid (0.3 g) were mixed with octadecene (5.15 g) in a round bottom flask (50 mL) under vigorous stirring. The reaction was performed under inert atmosphere. The temperature was increased at a constant heating rate (5 $^\circ$C per min) from 30 $^\circ$C to 320 $^\circ$C. The reaction was kept 1 h at this temperature and then cooled down to room temperature. 3 washing steps with ethanol/acetone (1/1) were conducted. Finally, the nanocubes were resuspended in toluene.
\vspace{0.5cm}

The chemical synthesis of PEGylated ligands and the subsequent ligand exchange procedure were reported previously in references \cite{nanocubes,nanocubes2}, respectively. In brief, a reaction solution was prepared in a round bottom flask with polyethylene glycol (3 g, PEG-3000), gallic acid (170 mg) and 4-dimethylaminopyridine (24 mg) in tetrahydrofuran (100 mL) and dichloromethane (10 mL). A solution of DCC (1 g) in tetrahydrofuran (10 mL) was added dropwise to the reaction solution under inert N$_2$ atmosphere. The mixture was stirred overnight at room temperature. The reaction mixture was filtered through a filter paper and the solvents were rotaevaporated. Solvents were supplied by Acros organics and the remaining reagents by Sigma Aldrich and used without further purification. The final solution contains a Fe concentration of 0.6 mg$\cdot$mL$^{-1}$ as determined with standard protocols by Inductively Coupled Plasma Hight Resolution Mass spectroscopy (ICPMD, NexION ICP-HRMS from Perkin-Elmer).
$^1$H nucler magnetic resonance (NMR Bruker Ascend 400 MHz spectometer) spectroscopy confirmed the desired product gallol-PEG-OH. $^1$H NMR (400 MHz, CDCl$_3$) $\delta$ (ppm): 7.22 (s,2H), 4.43-4.40 (m,2H), 3.85-3.45 (m,CH$_2$-PEG,-OH).
\vspace{0.5cm}

For the ligand exchange, in a glass vial a solution with nanocubes (1 mL) at concentration of 10 g$\cdot$L$^{-1}$ of Fe, we added the PEGylated ligand (1.0 mL) in a concentration of 0.1 \textsc{M} in CHCl$_3$. Triethylamine (50 $\mu$L) was added subsequently. The mixture was sonicated for 1 h and kept 4 h at 50 $^\circ$C. At this point, it was diluted with a mixture of toluene (5 mL), MilliQ water (5 mL) and acetone (10 mL). The nanoparticles were then transferred into the aqueous phase. After that, the aqueous phase was collected in a round-bottom flask and the residual organic solvents were rotaevaporated. The gallol-derived nanocubes were purified in centrifuge filters with a molecular weight cutoff of 100 kDa at 450 rcf. In each centrifugation step, the functionalized nanoparticles were resuspended in MilliQ water. The purification step was repeated several times until the filtered solution was clear. After the purification, the gallol-derived nanoparticles were resuspended in MilliQ water. Finally, to promote colloidal stability and monodispersivity, the suspension was centrifuged (150 rcf) for 5 min before being placed onto a permanent magnet (0.6 T) for 5 min previous to supernatant collection.

\subsection{Physicochemical characterization of iron oxide nanocubes. }
The UV-Vis-NIR spectrum was recorded in a UV-VIS-NIR Spectrophotometer (Jenway 6705) with a quartz tray having a light path of 1 cm. Nanocubes dispersion were used at a concentration of $\sim$0.04 mg$\cdot$mL$^{-1}$. The size distribution and $\zeta$-potential measurements of the gallol-derived nanocubes were performed on a Zetasizer Nano ZS90 (Malvern, USA). The nanoparticles were dispersed in MilliQ water at a concentration of 100 mg$\cdot$L$^{-1}$ of Fe. The measurements were done on a cell type ZEN0118-low volume disposable sizing cuvette, setting 2.42 as refractive index with 173$^\circ$ Backscatter (NIBS default) as angle of detection. The measurement duration was set as automatic and three as the number of measurements. TEM images were obtained on a FEI Tecnai G2 Twin microscope operated at an accelerating voltage of 100 kV. TEM samples were prepared by dropping a preconcentrated nanocubes suspension at $\sim$1 g Fe per liter on a carbon-coated copper grid and letting the solvent evaporate. The nanocube size were calculated by averaging a hundred of nanoparticles.

\subsection{Decoration of microgels with iron oxide nanocubes}

The decoration of the positively charged microgels with negatively charged iron oxide nanocubes was performed by adding the microgel dispersion (10 $\mu$L) to an Eppendorf with MilliQ water (1 mL), and three different volumes of iron oxide nanocubes dispersion: 40 $\mu$L, 20 $\mu$L, and 8 $\mu$L, to produce the $\mu gel_{High}$, $\mu gel_{Med}$, and $\mu gel_{Low}$ samples, respectively. The three Eppendorfs were placed in a shaker overnight. The dispersion was centrifuged in an Eppendorf (8000 rpm) for 15 min and the supernatant was removed and replaced by MilliQ water. The decorated microgels were characterized by TEM (Thermo Fisher Scientific TALOS F200X), obtaining also their $\mu_e$ and $d_H$ as described in the previous section.

\subsection{Optical tweezers trapping}
We used the NanoTracker-II optical tweezers (JPK-Bruker) equipped with a 1064 nm laser. The fluidic chambers were built by depositing two fine seams of vacuum seal grease (Korasilon paste, Kurt Obermeier GmbH) on a \#1.5 glass slide and covered with a \#1.5 cover slip. Next, the sample was deposited in between the two glasses by capillarity and the cell was sealed with vacuum seal grease. Laser powers were selected in the range 8-358 mW. The trap focus was allocated $\simeq$ 10 $\mu$m above the bottom glass of the cell to avoid spurious capture of microgels from the cone of light below the trap and to make hydordynamic interaction with the wall negligible \cite{Volpe_book}. The photons that are forward scattered by a trapped microgel carry information about its motion within the trap, and are detected by photodiodes to track the X, Y and Z time traces. The setup allows for an easy detection of multi-trapping events that are visualized both in the live video as a sudden brightening of the newly trapped microgel and as a step-wise increase in the time traces, coupled with a significant increase in the corner frequency $f_c$ obtained from the power spectral density (PSD) of the traces. If any of these features was detected, we discarded the experiment and looked for an alternative particle. For each trapped microgel, we calibrated the trap by performing a Stokes sinusoidal drag calibration \cite{tolic2006calibration,Volpe_book}. In this calibration method, the piezo-stage performed sinusoidal oscillation in the X-direction with an amplitude of $X_{stage}=$340 nm at $f_{stage}=$20 Hz, and the traces for the X, Y and Z coordinates in volts were recorded from the four quadrant photodiodes for X-Y, and a photodiode for the Z direction, at a rate of 50 kHz for 5 s. The standard procedure to analyse these data consists in the calculation of the power spectral density (PSD) of the time trace, which is composed by a Lorentzian function plus a peak corresponding to the forced oscillation of the trapped particle:

\begin{equation}\label{eq:PSD}
    PSD(f) = \frac{D}{2\pi^2(f^2+f_c^2)} + \frac{X_{stage}^2}{4(1+f_c^2/f_{stage}^2)}\delta(f-f_{stage}),
\end{equation}\\

where the first addend in the rhs comes from the Brownian motion of the particle in the trap, while the second one is due to the forced oscillation, with $\delta(x)$ the Dirac delta function of argument $x$\cite{tolic2006calibration}. Fitting the experimentally obtained PSD to Equation \ref{eq:PSD}, we obtain estimates for the corner frequency $f_c$, the calibration factor $S_{V\rightarrow m}\left[\frac{m}{V}\right]$, and the diffusion coefficient without any assumption on the temperature, viscosity of the media or the $d_H$ of the microgel (see Supplemental Material). This process was performed for each microgel and $P$ in triplicate to calculate averages and standard deviations. For the bare microgels, we could assume $\eta$ to be the viscosity of water at room temperature and obtain the value of the hydrodynamic diameter $d_H$.
\vspace{0.5cm}

Videos at a frame rate of 92 fps were also recorded to enable direct comparisons with the traces in volts stored during the tracking of the microgel with the Trackpy software \cite{Crocker1996methods}. As a result of the tracking, the centroid position in the XY plane and the radius of gyration of the brightness of the trapped microgel were obtained.
\vspace{0.5cm}

Finally, the electrophoretic mobility $\mu_{e,trap}$ measurements within the optical trap were performed by placing two gold nanowires electrodes of diameter 50 $\mu$m separated $d\simeq$0.5 mm \cite{Pesce_mu_e, Ma_mu_e}. Then, a voltage of $V_{pp}$=4 V was applied at a frequency of $f_{EF}=$20 kHz with a function generator (RSDG 2082X), while keeping the optical trap fixed with a microgel in between the two electrodes. By fitting the area of the peak that appears in the PSD at 20 kHz due to electrophoresis $W_{exp,EF}$, we calculated the value of $\mu_{e,trap}$ for each microgel at each $P$ value using Equation 5 \cite{Pesce_mu_e, Ma_mu_e}, which in analogy to the "Stokes calibration" does not depend on $T$, $\eta$, or the size of the trapped microgel. The expression relating optical tweezers parameters with electrophoretic mobility is

\begin{equation}
 |\mu_{e,trap}|= \frac{4\pi f_c }{E}\sqrt{W_{exp,EF}\left (  1+\frac{f_c^2}{f_{EF}^2} \right )},
\end{equation}
where the electric field $E$ is considered to be constant and given by $E=V_{pp}/2d$ (see Supplementary material for further details). Triplicate experiments were computed to obtained averages and standard deviations (see Figure S4b-c). 

\medskip
\textbf{Acknowledgements} \par 
 This work was supported by the projects EQC2018-004693-P, PID2021-127427NB-I00, PID2020-116615RA-I00, PID2020-118448RB-C21, and PID2019-105195RA-I00, and the grant IJC2018-035946-I funded by MCIN/AEI/ 10.13039/501100011033/ FEDER, UE, and by the project P18-FR-3583 and EMERGIA grant with reference EMC21\_00008 funded by Consejer\'ia de Universidad, Investigaci\'on e Innovaci\'on de la Junta de Andaluc\'ia. \\
 \dag M.A.F.R and S.O.B. contributed equally to this work.

\medskip

%
\bibliography{arxive_and_SI}

\clearpage
\newpage
\onecolumngrid
\begin{figure*}\centering
\textbf{Table of Contents}\\
\medskip
  \includegraphics{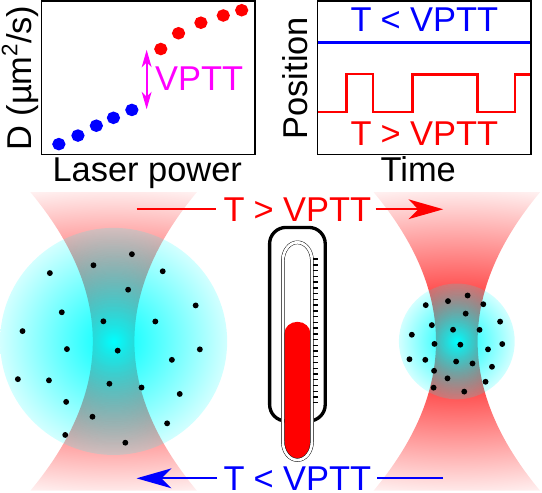}
  \medskip
  \caption{Self-heating microgels trapped in optical tweezers show discontinuous transitions and bistability when heated above the volume phase transition temperature. This new system can be used as local micro-heaters and micro-thermometers.}
\end{figure*}

\clearpage
\newpage

\renewcommand{\thefigure}{\textbf{S\arabic{figure}}}
\setcounter{figure}{0}    

\begin{figure}
\begin{center}
\LARGE{\textbf{Supplementary Information}}    
\end{center}\vspace{0.5cm}
  \includegraphics[width=0.9\linewidth]{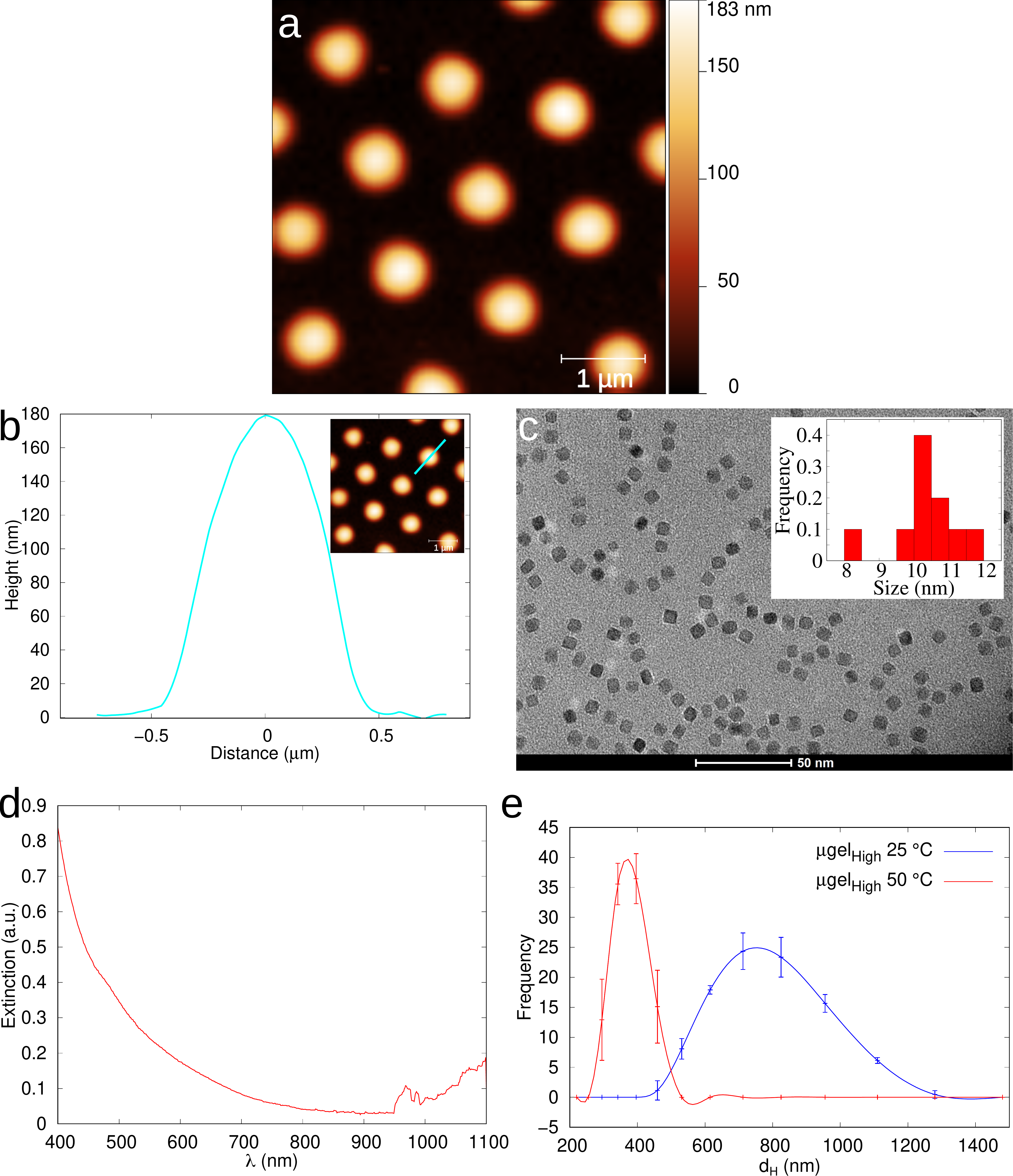}
  \caption{\textbf{(a)} AFM image of a microgel monolayer deposited at $\Pi$=5 mN$\cdot$m$^{-1}$. \textbf{(b)} Gaussian profile of the height of a microgel corresponding to the cyan section in the AFM image inset. \textbf{(c)} TEM image of the magnetite nanocubes. Inset with the distribution of diagonal lengths. \textbf{(d)} UV-vis-NIR extinction spectrum of the magnetic nanocubes in water. \textbf{(e)} Distribution of $d_H$ measured by DLS of microgels decorated with magnetite nanocubes $\mu gel_{High}$ in water.}
  \label{fig:S1}
\end{figure}

\clearpage
\section*{Stokes active calibration}
Calibration of the QPD signal to operative length units is required. The commercial optical tweezers provide with an automatic passive calibration procedure that adjust the PSD of a trapped particle and compare it with theoretical expressions. However, this method assumes constant local temperature $T_0$, viscosity and particle size, assumed to be spherical. As our study is focused on self-heating soft microparticles, none of those assumptions are valid, and an alternative calibration method is required. In this work, an active calibration method was selected. Within this approach, the voltage-to-length constant is derived from viscous drag (Stokes) measurements without any premise about the properties of the medium or the trapped particle \cite{Volpe_book}. To that aim, the uncalibrated particle’s temporal traces were recorded while applying a sinusoidal movement of amplitude $X_{stage}$ and frequency $f_{stage}$ described by:

\begin{equation}\tag{S1}
x(t)=X_{stage} \sin{(2\pi f_{stage} t)}.
\end{equation}

\begin{figure}
  \includegraphics[width=\linewidth]{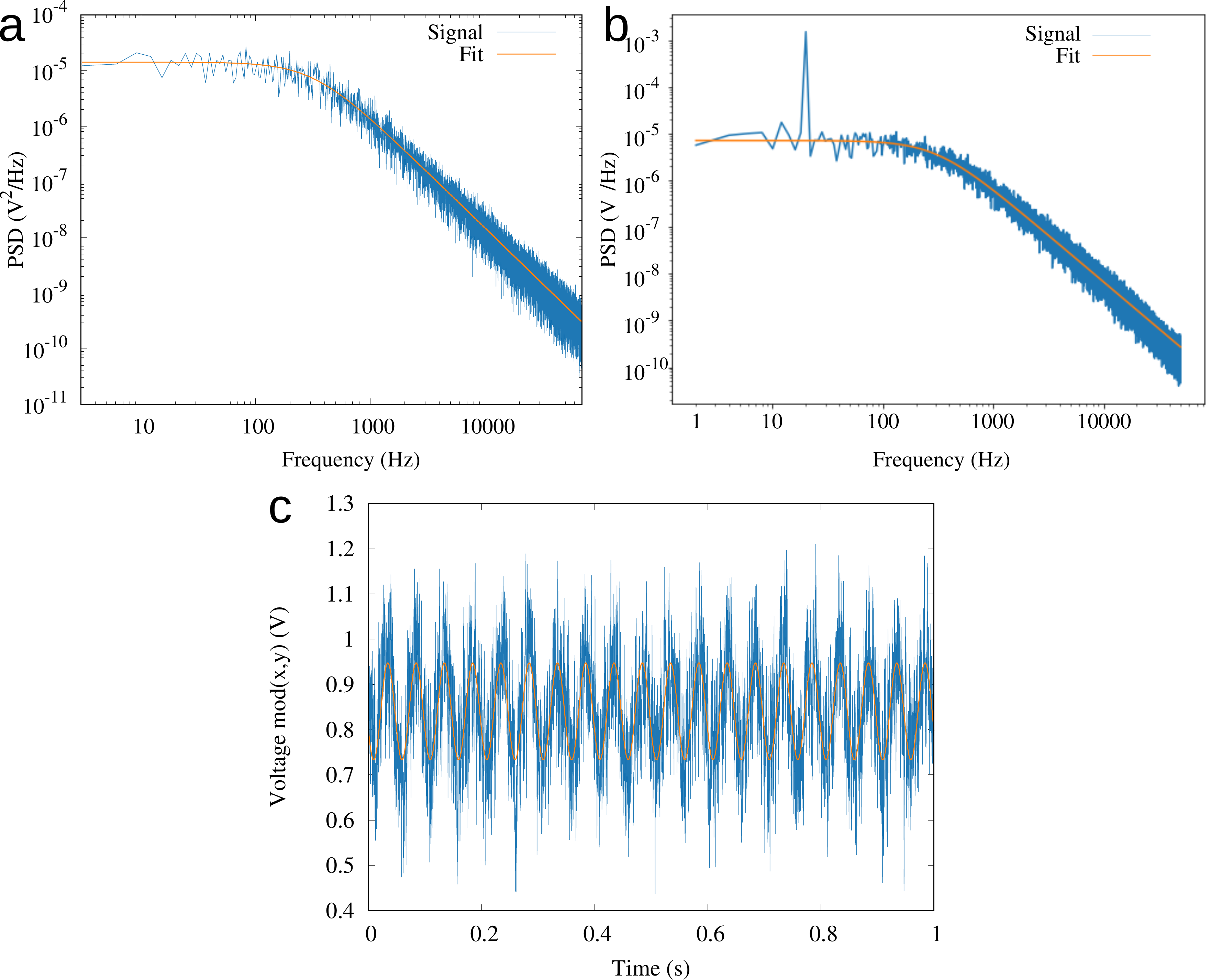}
  \caption{\textbf{(a)} Typical PSD of a trapped bare microgel ($P$=79 mW). \textbf{(b)} PSD of the same microgel in (a) but subjected to a sinusoidal drag of the stage of 340 nm amplitude and a frequency of 20 Hz, keeping the trap fixed. \textbf{(c)} X trace of the sinusoidal drag corresponding to the PSD in (b).}
  \label{fig:S2}
\end{figure}

The so-obtained PSD$_{exp,V}$ is formed by the superposition of the a peak corresponding to the sinusoidal trace over the Lorentzian curve at $f_{stage}$ (see \textbf{Figure \ref{fig:S2}}):

\begin{equation}\tag{S2}
    PSD_{exp}^{volts}(f)\left[\frac{V^2}{Hz}\right] = \frac{A}{(f^2+f_c^2)} + W_{exp}t_{msr}\delta_{Kronecker}(f-f_{stage}). 
\end{equation}\\

Experimentally, the measurement time $t_{msr}$ is finite, and therefore the Dirac delta in Equation 3 of the main text here becomes a Kronecker delta. Therefore, from a fitting to the experimental data we obtain the parameters $A$, $f_c$, and $W_{exp}$ without any assumption (see \textbf{Figure \ref{fig:S2}a}). Comparing with Equation 3 in the main text, we obtain the calibration factor:

\begin{equation}\tag{S3}
S_{V\rightarrow m}\left[\frac{m}{V}\right]=\sqrt{\frac{X_{stage}^2t_{msr}}{4W_{exp}(1+f_c^2/f_{stage}^2)}},
\end{equation}

and from it we can calibrate the PSD in length units:
\begin{equation}\tag{S4}
PSD_{exp}^{length}=S_{V\rightarrow m}^2\cdot PSD_{exp}^{volts} ,
\end{equation}

and a measurement of the diffusion coefficient:

\begin{equation}\tag{S5}
D=\frac{k_{B}T}{\gamma}=2\pi^2 A\cdot S_{V\rightarrow m}^2.
\end{equation}

\clearpage
\section*{Electrophoretic mobility of trapped microparticles}
In order to correlate the diffusion coefficient $D$ of the optically trapped microparticles with their electrophoretic mobility $\mu_{e,trap}$, we use the method described in the Experimental Section derived from \cite{Pesce_mu_e, Ma_mu_e}. The method is similar to the "Stokes active calibration" described before. For each trapped microparticle and laser power $P$ we measured by triplicate the microparticle temporal trace $x(t)$ while applying a sinusoidal electric field of amplitude $A_{EF}$ generated by a peak-to-peak voltage $V_{pp}=$4 V at $f_{EF}=$20 kHz. If the electric field $E$ is considered to be constant across the distance between electrodes $d$:

\begin{equation}\tag{S6}
E=\frac{V_{pp}}{2d}.
\end{equation}

Then, the expression for PSD$_{exp,EF}$ can be approximated by: 
\begin{equation}\tag{S7}
    PSD_{exp,EF}(f) = \frac{A_{exp}^{length}}{f^2+f_c^2} + \frac{A_{EF}^{length}}{4(1+f_c^2/f_{EF}^2)}\delta(f-f_{EF}),
\end{equation}\vspace{0.1cm}

which is quite similar to Equation S2, just changing the origin of the sinusoidal movement and taking the values in length units as obtained by the calibration procedure described above.\\

In equilibrium, the force due to the electric field $F_{EF}$, given by:
\begin{equation}\tag{S8}
F_{EF}=Q_{eff}E,
\end{equation}\\

is counterbalanced by the optical restoring force $F_{trap}$ determined by the stiffness $\kappa$:

\begin{equation}\tag{S9}
F_{trap}=\kappa A_{EF}^{length}.
\end{equation}\\

In Equation S8 $Q_{eff}$ is the effective charge of the trapped microparticle, related to the electophoretic mobility by:

\begin{equation}\tag{S10}
Q_{eff}= |\mu_{e,trap}|\gamma.
\end{equation}\\

It is straightforward to show that:

\begin{equation}\tag{S11}
A_{EF}^{length}=\frac{Q_{eff}E}{\kappa}.
\end{equation}\\

While the particle is moving, $F_{EF}$ must be counterbalanced by the drag force $F_{drag}$:
\begin{equation}\tag{S12}
F_{drag}=|\mu_{e,trap}|\gamma E.
\end{equation}\\

and an operative form relating $A_{EF}^{length}$ with $\mu_{e,trap}$ \cite{Ma_mu_e} can be obtained:\\

\begin{equation}\tag{S13}
    A_{EF}^{length}=\frac{|\mu_{e,trap}|\gamma E}{\kappa}=\frac{|\mu_{e,trap}|E}{2\pi f_c}.
\end{equation}\\

Similarly to the active calibration, we obtain the area under the experimental peak $W_{exp,EF}$ and compared to the theoretical one $W_{th,EF}$ given by:\\

\begin{equation}\tag{S14}
W_{th,EF}=\int_{0}^{f_{Nyquist}}PSD_{EF}(f)df=\frac{(A_{EF}^{length})^2}{4(1+f_c^2/f_{stage}^2)}=\frac{\mu_{e,trap}^2 E^2}{16\pi^2 f_c^2 (1+f_c^2/f_{stage}^2)}.
\end{equation}\\
 Finally, we can obtain the absolute value of $\mu_{e,trap}$ as reported in Equation 5 in the main text, reproduced here:\\ 

\begin{equation}\tag{S15}
 |\mu_{e,trap}|= \frac{4\pi f_c }{E}\sqrt{W_{exp,EF}\left (  1+\frac{f_c^2}{f_{EF}^2} \right )}.
\end{equation}

It must be emphasized that no assumption for $T$, $\eta$, nor the size or shape of the trapped microparticle have been assumed during the derivation. As discussed in the manuscript, these values, although in the correct units and order of magnitude, are difficult to compare to the ones obtained by Laser Doppler Micro-electrophoresis in Figure 1d. In the latter, a 10 times higher electric field is applied, and the velocity is obtained by Doppler shift from the laser pointed at the sample. In any case, we use them here as an independent measurement to find the discontinuous transition of each trapped self-heating microgel.\\

\begin{figure}
  \includegraphics[width=\linewidth]{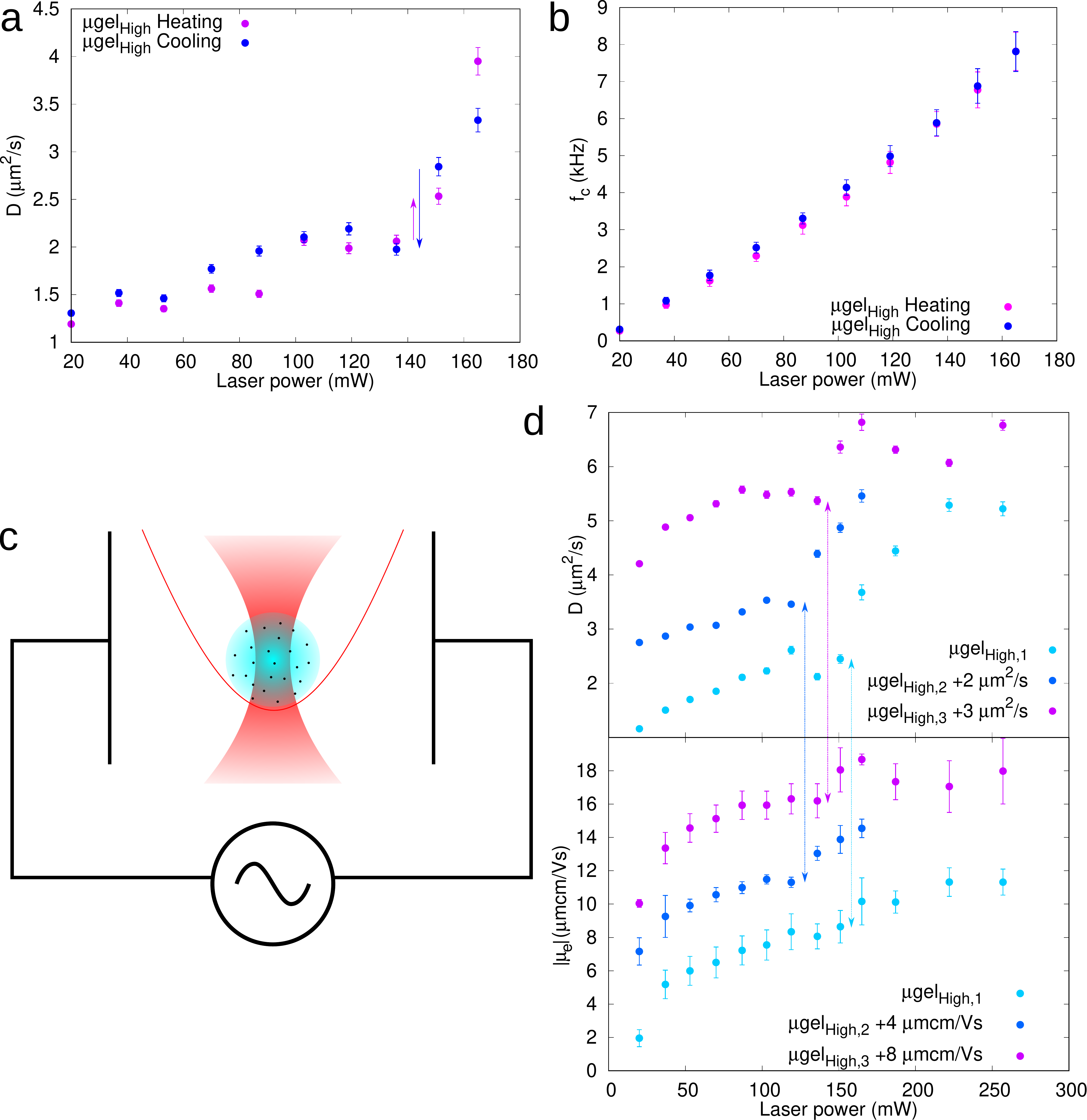}
  \caption{\textbf{(a)} $D$ of one microgel as a function of $P$ upon heating and then cooling, to show the reversibility of the process. Arrows depict the discontinuous transitions. \textbf{(b)} $f_c$ corresponding to the curves in (a). \textbf{(c)} Scheme of the measurement of $|\mu_{e,trap}|$. \textbf{(d)} Figure 2b (diffusion coefficients $D$) plotted next to the independent measurements of $|\mu_{e,trap}|$ for each microgel. Arrows pointing to the discontinuous transitions.}
  \label{fig:S4}
\end{figure}

\begin{figure}
  \includegraphics[width=\linewidth]{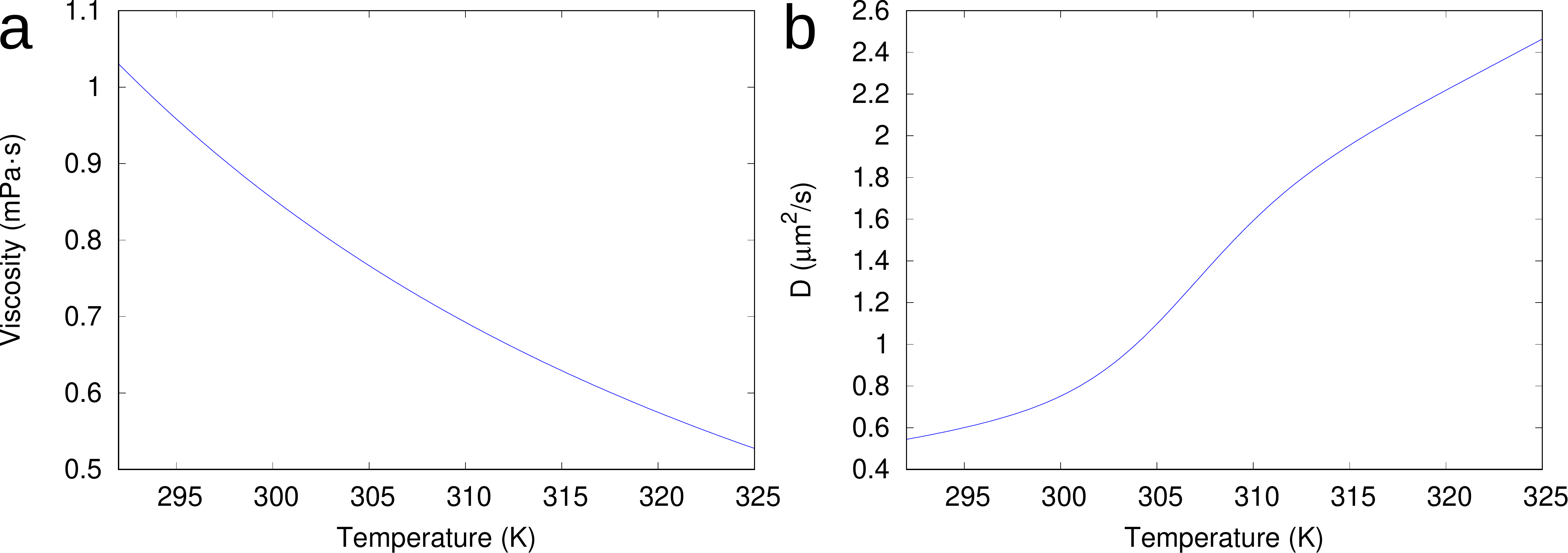}
  \caption{\textbf{(a)} Viscosity $\eta$ of water as a function of $T$ \cite{eta}, that can be approximated by the equation $\eta(T)=2.414\cdot10^{-5}\cdot10^{247.8 \rm{K}/(T-140 \rm{K})}\,\rm Pa\cdot s$ \textbf{(b)} Diffusion coefficient $D$ expected for a single $\mu gel_{High}$ as a function of $T$ as obtained from Equation 1 and 2.}
  \label{fig:S3}
\end{figure}

\begin{figure}
\centering
  \includegraphics[width=0.9\linewidth]{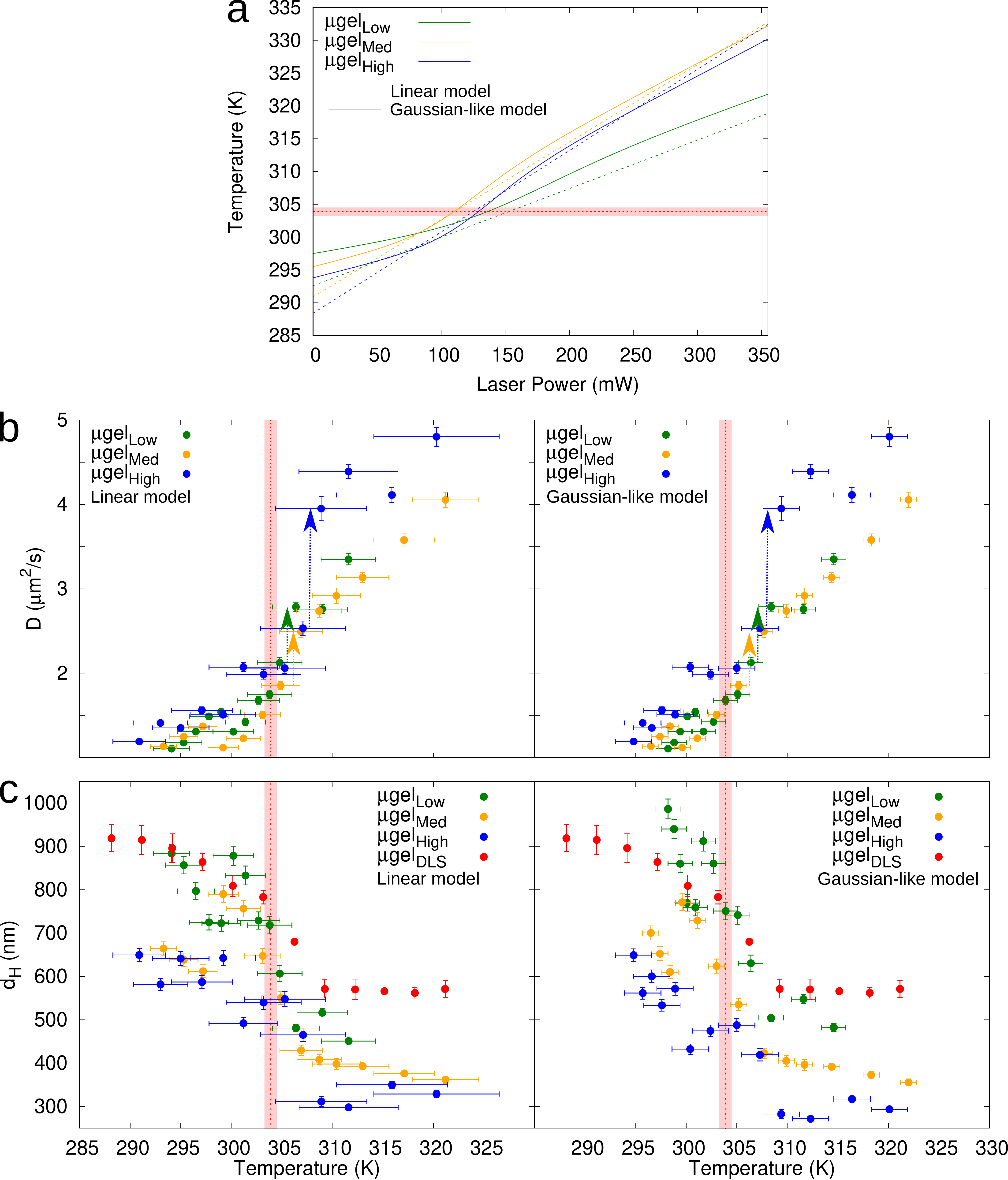}
  \caption{\textbf{(a)} $T(P)$ for the linear (dashed lines and data points with errors) and Gaussian-like (solid lines) fittings in Figure 2c and Table 1. \textbf{(b)} $D$ from Figure 3a as a function of $T$, calibrated with the linear and Gaussian-like fitting. \textbf{(c)} $d_H$ of the microgels as a function of $T$ from (b). Bare microgels characterized by DLS were added for comparison. The $VPTT=303.9$ K is represented by red dashed lines. Light red shaded regions stand for the $\pm0.6$ K interval.}
  \label{fig:S5}
\end{figure}

\begin{figure}
\centering
  \includegraphics[width=0.9\linewidth]{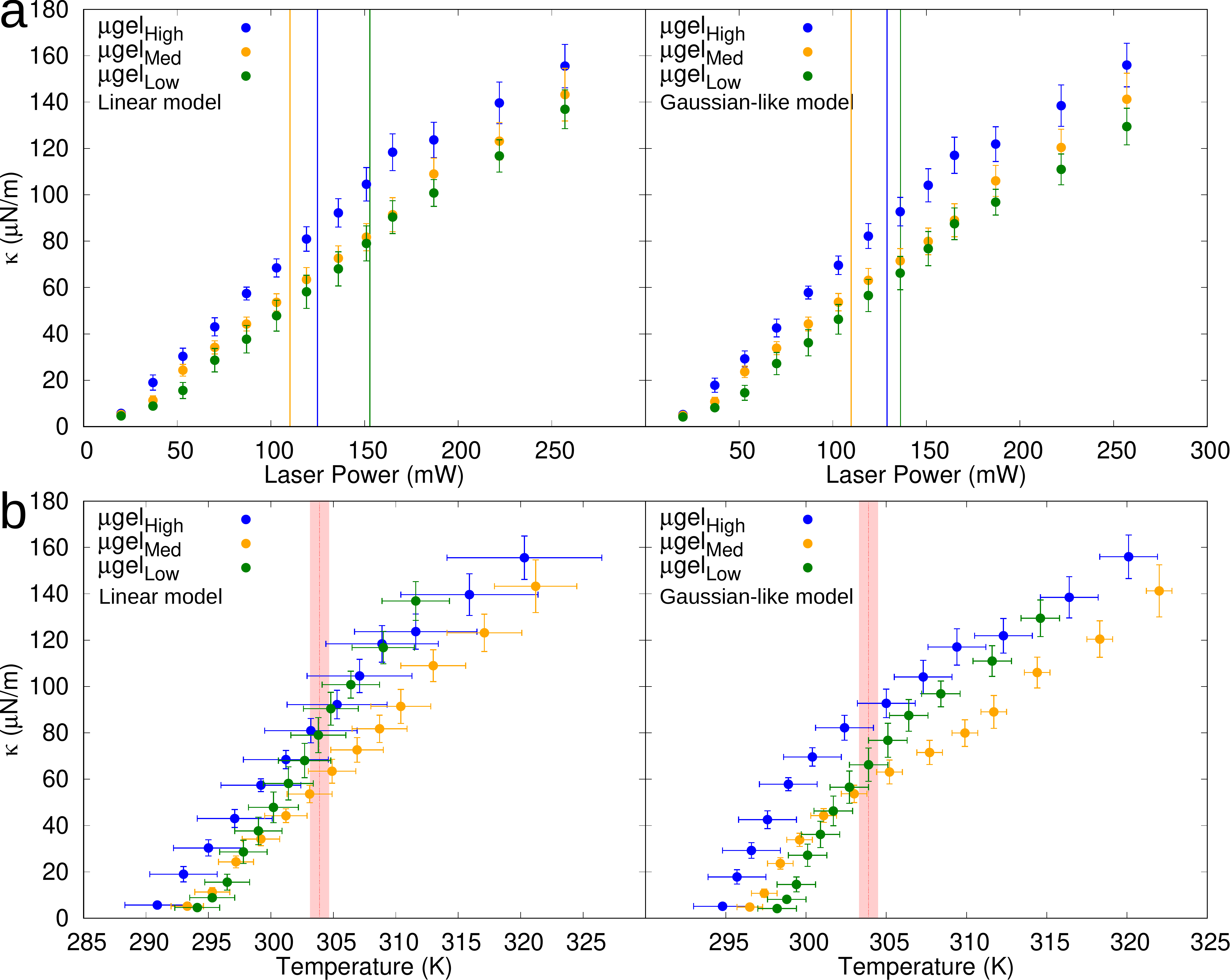}
  \caption{Stiffness of the trap $\kappa= 2\pi f_c \gamma_{\rm SHBM} = 6\pi^2 f_c \phi \eta(T) d_H(T)$ as a function of \textbf{(a)} $P$ and \textbf{(b)} $T$ for the linear and Gaussian-like temperature calibrations. The $VPTT=303.9$ K is represented by solid lines in (a) and red dashed lines in (b). Light red shaded regions stand for the $\pm0.6$ K interval.}
  \label{fig:S6}
\end{figure}

\begin{figure}
  \includegraphics[width=\linewidth]{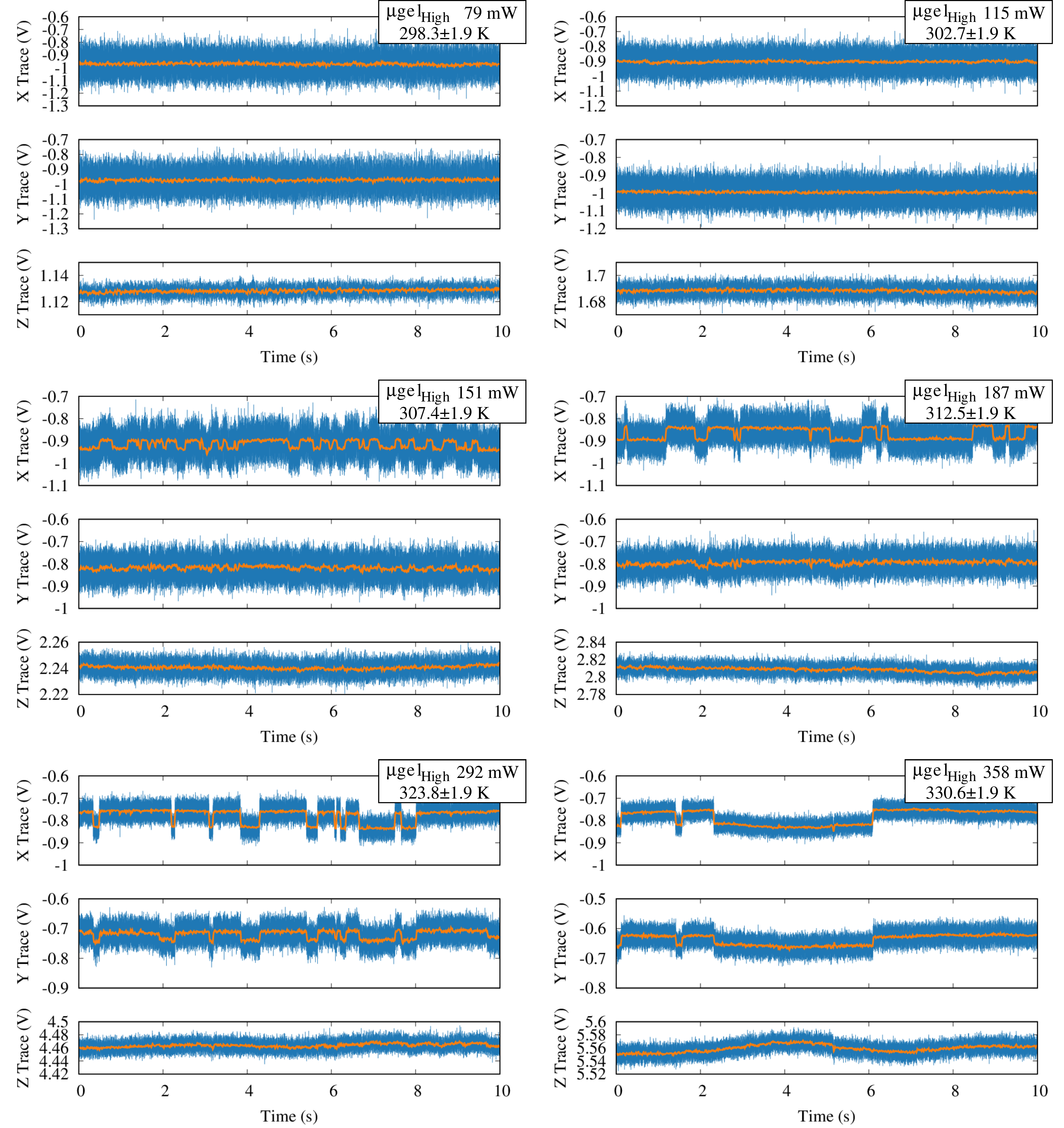}
  \caption{\textbf{(a)} Traces for the same microgel as in Figure 4a at different $P$ and $T$ values. The Gaussian fitting was used to estimate $T$.}
  \label{fig:S7}
\end{figure}

\begin{figure}
  \includegraphics[width=\linewidth]{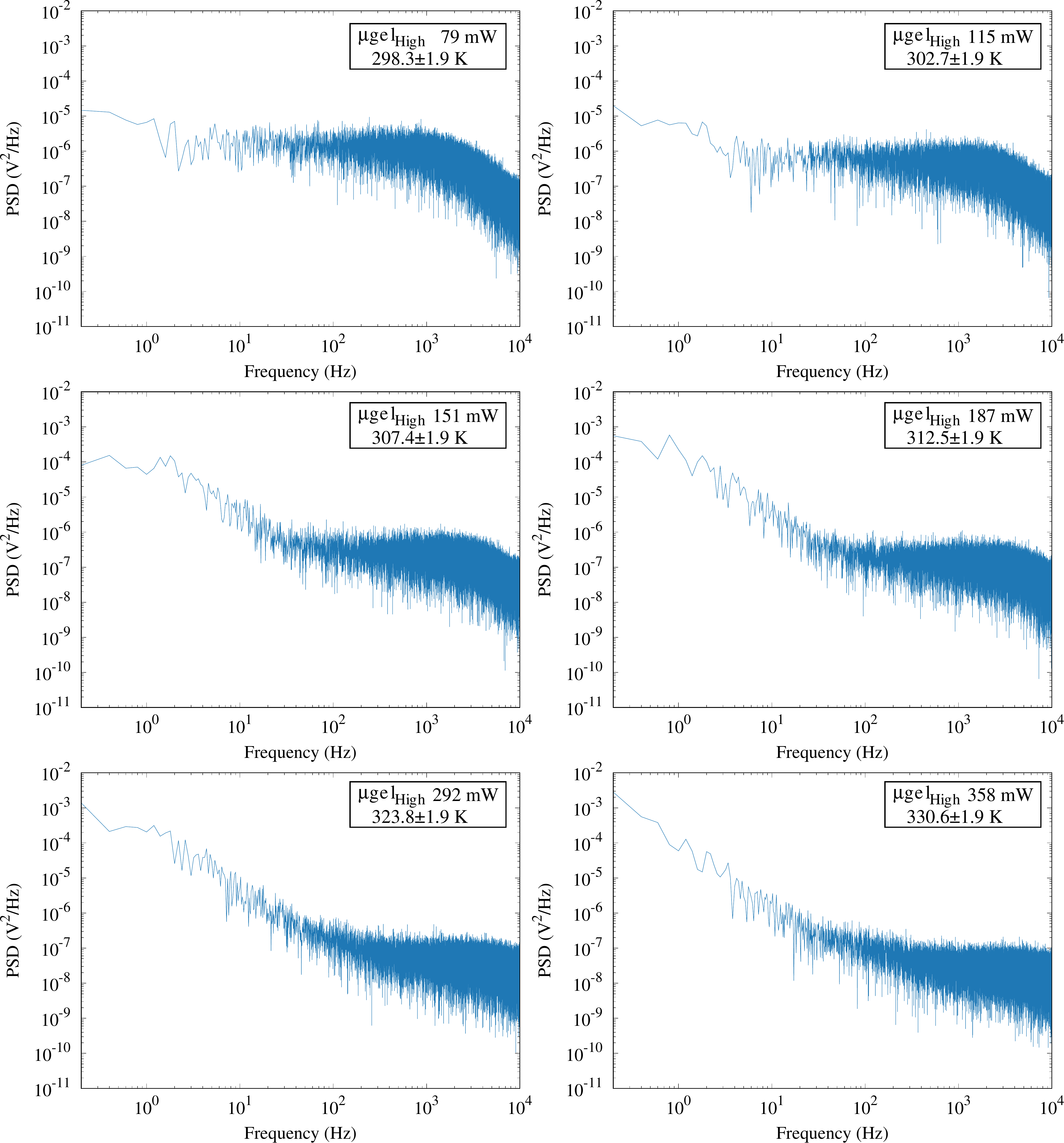}
  \caption{PSDs corresponding to the measurements in Figure S7.}
  \label{fig:S8}
\end{figure}

\begin{figure}\centering
  \includegraphics[width=0.9\linewidth]{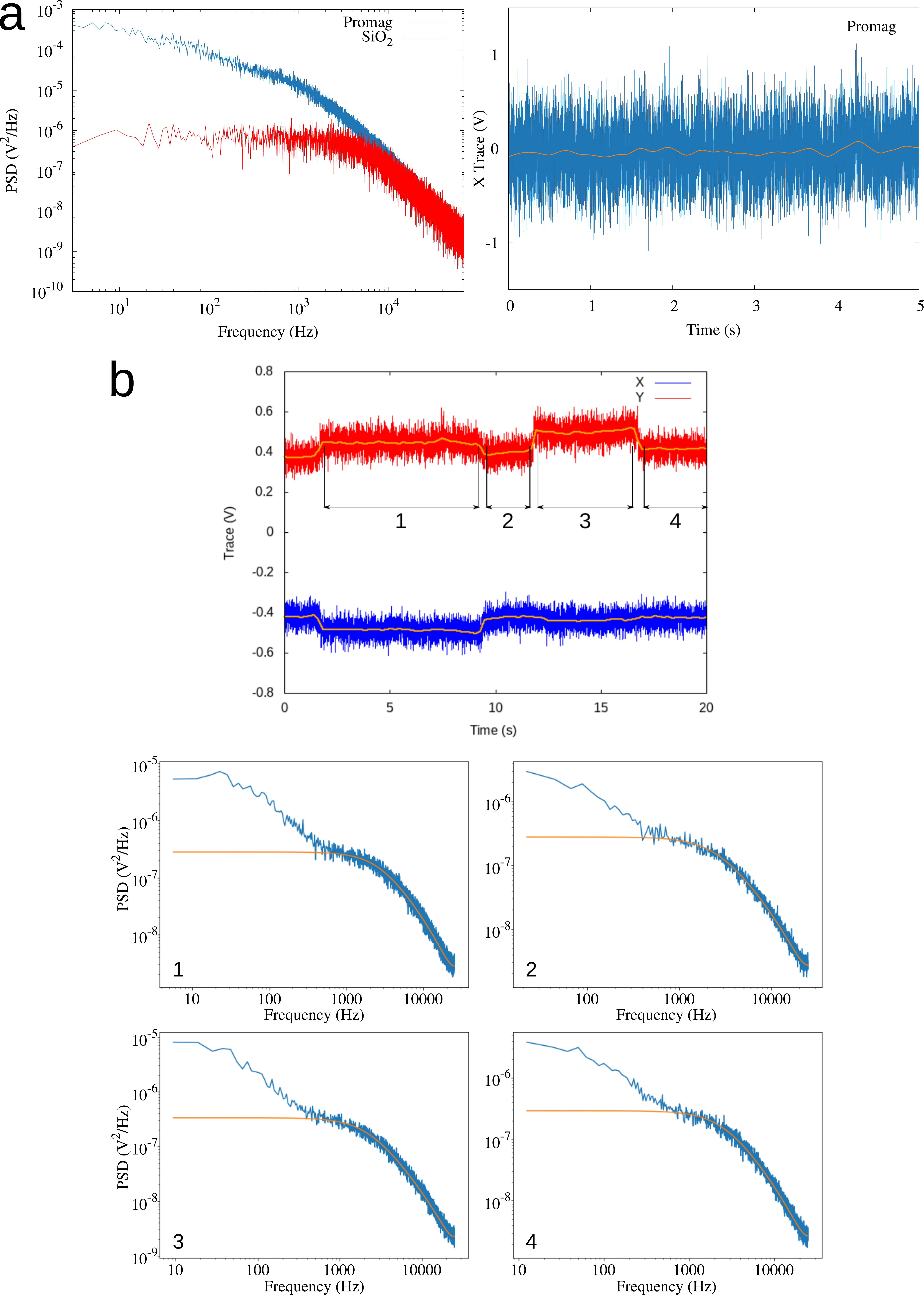}
  \caption{\textbf{(a)} PSDs of the X trace of a non-absorbing 1 $\mu$m hard silica particle ($P$=358 mW) and a self-heating hard Promag microparticle ($P$=28 mW), and X trace showing no bistability for the latter (smooth curve in orange). Promag microparticles are polymer-based magnetite-loaded micro-spheres of 1 $\mu$m. \textbf{(b)} PSDs of the Y trace in volts corresponding to the different domains in the bistability shown in Figure 3c-d.}
  \label{fig:S9}
\end{figure}

\clearpage
\section*{Scattering force vs bulk elasticity}
We can estimate the scattering force provided by the iron oxide nanocubes within the decorated microgel \cite{Pesce2020} as:

\begin{equation}\tag{S16}
F_{scattering}=N\frac{n_m \sigma_{ext}}{c} I_0 e^{-2 \left(\frac{\rho}{w}\right)^2}
\end{equation}

where $N$ is the number of iron oxide nanocubes per microgel. This was estimated to be $N\simeq 1600$ in Figure 1b (tracked by imageJ) for a $\mu gel_{High}$. The refractive index of the medium is the one of water $n_m=1.33$, $c$ is the speed of light, $\sigma_{ext}\simeq2.5$ nm$^{2}$ for quite similar iron oxide nanocubes \cite{magnetite_absorption} at 1064 nm, and $\rho$ is the distance of the nanocubes to the center of the trap. $I_0=P/(\pi w^2)$, where $P$ is the laser power and $w=280$ nm is the beam radius of our laser. Finally, we used $\rho=d_{H,collapsed}\cdot r/2=366/4$ nm, where $r\simeq0.5$ accounted for the position of the iron oxide nanocubes within the microgel as discussed in the main text.
\vspace{0.5 cm}

Further, we estimate the force needed to deform the microgel by a relative volume $\Delta V/V$ using the bulk compressibility modulus $K$, reported to be $\simeq 1$  kPa above the $VPTT$ for very similar microgels in size and composition, measured by AFM \cite{Bulk_modulus}. Similar results were reported from osmotic pressure measurements \cite{Bulk_modulus2}. The expression of $K$ is the following:

\begin{equation}\tag{S17}
K=-V\frac{dP}{dV}\simeq\frac{\Delta P}{\frac{\Delta V}{V}}
\end{equation}

In Equation S18, the pressure $\Delta P$ can be seen as a force $F$ on the decorated surface of the particle $\Delta P = F_{deformation}/(4\pi \rho^2)$, resulting in:

\begin{equation}\tag{S18}
F_{deformation}=4\pi \left(\frac{d_{H,collapsed}\cdot r}{2}\right)^2 K \frac{\Delta V}{V}
\end{equation}

We plotted Equation S16 and S18 in \textbf{Figure S10} as a function of $P$ and $\Delta V/V$ for the $F_{scattering}$ and $F_{deformation}$, respectively, showing that $F_{scattering}>F_{deformation}$. This result indicates that a deformation of the microgel along the propagation direction of the beam is feasible.

\begin{figure}[h]\centering
  \includegraphics[width=0.6\linewidth]{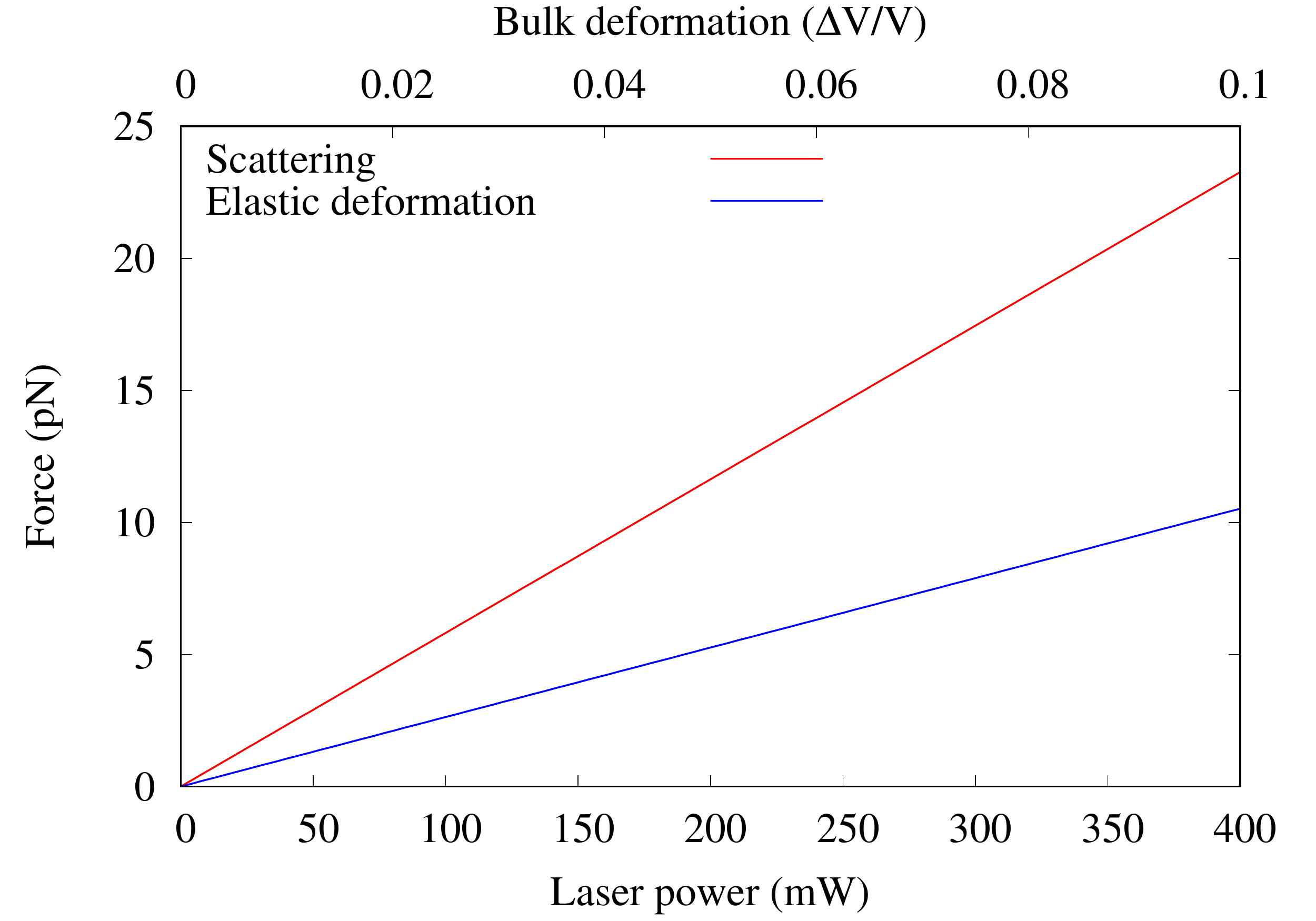}
  \caption{Comparison between the scattering force as a function of laser power $P$ and the force required to deform the microgel as a function of the deformation $\Delta V/V$.}
  \label{fig:S10}
\end{figure}

\clearpage
\textbf{Movie S1.} Movie acquired at 92 fps for the trapped $\mu gel_{Med}$ at $P$=358 mW used to track the position and radius of gyration of the brightness of the microgel in Figure 3d.


\end{document}